\documentclass[11pt,epsf]{article}
\usepackage{amsmath}
\usepackage{amsfonts}
\usepackage{amssymb}
\usepackage{graphicx}
\usepackage{color}
\usepackage{comment}

\newtheorem{lemma}{Lemma}
\newcommand{\dfn}{\stackrel{\triangle}{=}}
\newcommand {\exe} {\stackrel{\cdot} {=}}
\newcommand {\lexe} {\stackrel{\cdot} {\le}}
\newcommand{\eqa}{\stackrel{\mbox{\tiny (a)}}{=}}
\newcommand{\eqb}{\stackrel{\mbox{\tiny (b)}}{=}}

\newcommand {\reals} {{\rm I\!R}}

\newcommand {\btheta} {\mbox{\boldmath $\theta$}}
\newcommand {\bphi} {\mbox{\boldmath $\phi$}}
\newcommand {\bGamma} {\mbox{\boldmath $\Gamma$}}
\newcommand {\balpha} {\mbox{\boldmath $\alpha$}}

\newcommand {\bc} {\mbox{\boldmath $c$}}

\newcommand {\bx} {\mbox{\boldmath $x$}}
\newcommand {\by} {\mbox{\boldmath $y$}}

\newcommand {\bE} {\mbox{\boldmath $E$}}

\newcommand {\bI} {\mbox{\boldmath $I$}}

\newcommand {\bX} {\mbox{\boldmath $X$}}
\newcommand {\bY} {\mbox{\boldmath $Y$}}
\newcommand {\bZ} {\mbox{\boldmath $Z$}}
\newcommand{\calA}{{\cal A}}
\newcommand{\calB}{{\cal B}}

\newcommand{\calI}{{\cal I}}

\newcommand{\calN}{{\cal N}}

\newcommand{\calS}{{\cal S}}

\newcommand{\calX}{{\cal X}}

\begin{document}

\thispagestyle{empty}
\setcounter{page}{1}
\setlength{\baselineskip}{1.5\baselineskip}

\title{Volume-Based Lower Bounds to the Capacity of the Gaussian Channel Under
Pointwise Additive Input Constraints}


\author{Neri Merhav 
and Shlomo Shamai (Shitz)
\\ \\
The Andrew \& Erna Viterbi Faculty of Electrical and Computer Engineering\\
Technion - Israel Institute of Technology \\
Haifa 3200003, ISRAEL}
\maketitle

\begin{abstract}
We present a family of relatively simple and unified lower bounds on the 
capacity of the Gaussian channel under a set of pointwise additive input 
constraints. Specifically, the admissible channel input vectors
$\bx = (x_1, \ldots, x_n)$
must satisfy $k$ additive cost constraints of the form
$\sum_{i=1}^n \phi_j(x_i) \le n \Gamma_j$, 
$j = 1,2,\ldots,k$,
which are enforced pointwise for every $\bx$, rather than merely in 
expectation.
More generally, we also consider cost functions that depend on a sliding 
window of fixed length $m$, namely,
$\sum_{i=m}^n \phi_j(x_i, x_{i-1}, \ldots, x_{i-m+1}) \le n \Gamma_j$, 
$j = 1,2,\ldots,k$,
a formulation that naturally accommodates correlation constraints as well as a 
broad range of other constraints of practical relevance.

We propose two classes of lower bounds, derived by two methodologies that 
both rely on the exact evaluation of the volume exponent associated with the set 
of input vectors satisfying the given constraints. This evaluation exploits extensions of the method 
of types to continuous alphabets, the saddle-point method of integration, 
and basic tools from large deviations theory.
The first class of bounds is obtained via the entropy power inequality (EPI), and therefore applies 
exclusively to continuous-valued inputs. The second class, by contrast, is
more general, and it applies 
to discrete input alphabets as well. It is based on a direct
manipulation of mutual information, and it yields stronger and tighter bounds, though at the cost of greater 
technical complexity. Numerical examples illustrating both types of bounds are provided, and several 
extensions and refinements are also discussed.\\

{\bf Index Terms:} Gaussian channel, channel capacity, entropy power
inequality, peak-power constraint, volume exponent, saddle-point method, large deviations. 
\end{abstract}

\clearpage

\section{Introduction}

Input-constrained Gaussian channels have been a central topic 
of study since the earliest days of information theory \cite{SW49}, and they continue to 
be addressed in numerous papers and textbooks (see, e.g., \cite{CT06} and references therein, as well as those cited here).
The most common constraint is the average power constraint, which reflects the physical power budget of the 
transmitter and leads to the classical expression for the Gaussian channel capacity \cite{SW49}. 
Beyond this standard setting, additional constraints capture further limitations of practical signaling. 
For instance, the peak-power constraint \cite{Smith71} has yielded fundamental
insights, most notably, the result that 
the capacity-achieving input distribution for the discrete-time memoryless Gaussian channel is discrete with finite support.

In practice, discrete input constellations are employed in virtually all communication systems, 
regardless of whether they coincide with the true capacity-achieving distribution 
(which is Gaussian under an average-power-only constraint). 
Discrete inputs are also used to quantify the performance loss due to practical signaling restrictions, 
even in cases where Gaussian inputs remain theoretically optimal \cite{OW90}. From this perspective, the theoretical 
results of \cite{Smith71} carry clear practical significance.
Moreover, constrained Gaussian input models play a central role in the study of optical 
communication channels, where the key additional constraint is non-negativity
of the input signal, imposed by intensity modulation \cite{CRA22, LMW09, SS10}. 
Similar constrained models have also been considered in other contexts (see, e.g., \cite{FAF18, SBD95}).

Discrete-time Gaussian channels with filtered inputs, which induce inter-symbol interference (ISI), provide 
both a natural and practically significant model for communication systems under realistic constraints. 
These channels have been investigated for many decades, leading to a rich body
of results and a variety of bounds, 
some of which are formulated in terms of equivalent scalar Gaussian channels \cite{HM88, SOW91}. 
Similarly, continuous-time filtered Gaussian channels subject to average- and 
peak-power constraints have also been extensively studied, motivated by 
the need to accurately model practical systems (see, e.g., \cite{PS24} and references therein).

The exact evaluation of capacity under practical constraints, 
such as peak-power limitations, is notoriously difficult and typically requires intricate 
numerical procedures. This challenge motivates the study of simple lower and upper bounds on capacity, 
a line of work well represented in the literature (see \cite{TKB17} and references therein). 
Capacity bounds under peak-power and related practical constraints have also been extensively investigated 
in broader settings, including vector Gaussian channels, multiple-input multiple-output 
(MIMO) Gaussian channels, and related models \cite{DAPS19, DGPS19, DYPS20} and
references therein.

As evidenced by the vast literature on discrete-time Gaussian channel models, exact capacity characterizations 
are rarely attainable. This scarcity strongly motivates the development of relatively simple lower bounds on Gaussian 
channel capacity under a broad class of constraints, extending beyond the classical average- and 
peak-power limitations. Examples include, for instance, moment-based constraints studied in \cite{EMM23, GTS25}.

In this work, we present a family of relatively simple and unified lower bounds to the
capacity of the Gaussian channel subject to a set of point-wise additive
constraints. Specifically, the allowable channel input vectors,
$\bx=(x_1,\ldots,x_n)$, must comply with a set of $k$ additive cost constraints of the form,
\begin{equation} 
\label{memorylessconstraints}
\sum_{i=1}^n\phi_j(x_i)\le n\Gamma_j,~~~~~j=1,2,\ldots,k,
\end{equation} 
where $\{\phi_j(\cdot),~j=1,2,\ldots,k\}$ are given cost functions and
$\{\Gamma_j,~j=1,2,\ldots,k\}$ are given numbers. Note that these constraints are
imposed point-wise, for every $\bx$, and not merely in expectation. The most
common example is, of course, the average power constraint, corresponding to
$k=1$ and $\phi_1(x)=x^2$ and $\Gamma_1=P$. If, in addition, one wishes to
add, for example,
a peak-power constraint, this can be addressed in this
framework, by defining $k=2$, $\phi_1(x)=x^2$, and $\phi_2(x)$ being defined
as $\phi_2(x)=0$ for $|x|\le A$ and
$\phi_2(x)=\infty$ for $|x|>A$.
More generally, we can also allow cost functions that depend on a sliding-window of a
fixed size, $m$, i.e.,
\begin{equation} 
\label{constraintswithmemory}
\sum_{i=m}^n\phi_j(x_i,x_{i-1},\ldots,x_{i-m+1})\le n\Gamma_j,~~~~~j=1,2,\ldots,k.
\end{equation} 
These are useful to impose e.g., correlation constraints, where
$\phi_j(x_i,x_{i-\ell})=x_ix_{i-\ell}$, as well as a variety of many other
practically relevant constraints, as will be discussed in the sequel. We
henceforth refer to the constraints of the form
(\ref{memorylessconstraints}) as {\em memoryless constraints}, and to
constraints of the form (\ref{constraintswithmemory}) for $m\ge 2$, as 
{\em constraints with memory}, or as {\em sliding-window constraints}.

The proposed lower bounds,
in their basic forms, depend on $k$ parameters 
over which an expression should
be minimized. In other words, the number of parameters to be optimized is equal to the number
of constraints. In some of the
more sophisticated versions of the proposed lower bounds, there are additional
parameters, but these are parameters for maximization, and so, the
maximization is not mandatory as an arbitrary choice of the values of those additional
parameters are adequate for the purpose of obtaining a valid lower
bound.

We propose two classes of lower bounds, which are derived by two methodologies that
are based on exact evaluation of the volume exponent associated with the set
$\calS_n\subset\reals^n$ of input vectors that satisfy the constraints
(\ref{memorylessconstraints}), or more generally,
(\ref{constraintswithmemory}). This evaluation of the volume exponent is based
on extensions of the method of types to continuous alphabets and on
saddle-point integration methods \cite[Chapters 2 and 3]{MW25} (see also
\cite{deBruijn81}) as well as elementary large deviations theory \cite{DZ98}. Of course, for
a finite channel input alphabet, the ordinary method of types can be used
instead, The first class of bounds is based on 
the entropy-power inequality (EPI) (see, e.g., \cite{CT06}) and therefore applies to
continuous-valued inputs only, and the second class applies also to
discrete-alphabet inputs. The second class of bounds, which builds on direct
evaluation of the mutual information,
is stronger and tighter, but somewhat
more complicated. Several extensions and modifications are also discussed.

It is important to emphasize that we do not claim that our bounds are tighter
than all bounds reported in the literature for each and every specific model. Our
contribution lies in proposing systematic methodologies for deriving good lower bounds
in a rather general framework of channel input constraints, including
sliding-window constraints (constraints with memory), which are not trivial to
handle, in general.

The outline of the remaining part of this article is as follows. In
Section \ref{ncso}, we establish our notation conventions (Subsection
\ref{nc}), provide a formal
description of the setting (Subsection \ref{setup}), and specify our
objective (Subsection \ref{objective}). In Section \ref{epi}, we present the
basic EPI lower bound (Subsection \ref{basicepi}), provide the volume
exponent formula, first, for memoryless constraints (Subsection
\ref{volume-exponent}), demonstrate it in a couple of examples (Subsection
\ref{someexamples}), and finally, extend the scope to constraints with memory
(Subsection \ref{slidingwindowconstraints}) along with some additional
examples (Subsection \ref{moreexamples}).
In Section \ref{directmanipulation}, we derive alternative lower bounds by
direct manipulation of the mutual information where the channel input
distribution is set to be uniform across the set of input vectors that comply
with all constraints. We do this mostly for memoryless constraints (Subsection
\ref{memorylessconstraints2}), but we also
outline the basics of a possible derivation for constraints with memory
(Subsection \ref{constraintswithmemory2}).
For memoryless constraints, we also discuss and demonstrate a possible further improvement based on
a certain parametric family of non-uniform input distributions. Finally, in Section \ref{conclusion}, we
summarize and conclude this work, along with an outlook for future work.

\section{Notation Conventions, Setup, and Objective}
\label{ncso}

\subsection{Notation Conventions}
\label{nc}

Throughout this paper, random variables are denoted by capital letters, 
their realizations -- by the corresponding lowercase letters, and their
alphabets -- by calligraphic letters. 
Random vectors and their realizations are denoted by boldface capital and lowercase letters, respectively, 
with their alphabets expressed as Cartesian powers of the underlying single-letter alphabet. 
For example, the random vector $\bX=(X_1,\ldots,X_n)$
may take a realization $\bx=(x_1,\ldots,x_n)$ in $\calX^n$, the $n$th power of
the single-letter alphabet, $\calX$.
For two positive integers $i$ and $j$, with $i< j$,
we use the shorthand notation $x_i^j$ to denote $(x_i,x_{i+1},\ldots,x_j)$
with the analogous convention for random variables, e.g.,
$X_i^j=(X_i,X_{i+1},\ldots,X_j)$. When $i=1$, the subscript will be omitted,
namely, $x^j$ and $X^j$ will stand for $x_1^j$ and $X_1^j$, respectively.

Sources and channels will be denoted by 
the letters $p$, $f$, and $q$,
subscripted by the names of the relevant random variables or vectors, including conditionings when appropriate, 
following standard conventions (e.g., $q_X$, $p_{Y|X}$, etc.).
When no ambiguity arises, these subscripts will be omitted.
The probability of an event $\calA$ will be
denoted by $\mbox{Pr}\{\calA\}$.
The expectation operator with respect to (w.r.t.) a probability distribution 
$p$ will be written as $\bE_p\{\cdot\}$,
with the subscript dropped whenever the distribution is clear from context.
If the underlying distribution depends on a parameter $\btheta$, we will
denote the expectation by $\bE_{\btheta}\{\cdot\}$. Likewise,
$\mbox{Pr}_{\btheta}\{\calA\}$ will denote probability w.r.t.\ the source
parameterized by $\btheta$.
Information measures follow the conventional notation of information theory: for example, 
$h(\bY)$ is the differential entropy of $\bY$,
$h(\bY|\bX)$ the conditional differential entropy of 
$\bY$ given $\bX$,
$I(\bX;\bY)$ is the mutual information between 
$\bX$ and $\bY$, etc.
Finally, for a probability function 
$q(\bx)$ (a probability mass function if 
$\bx$ is discrete, or a probability density function if it is continuous), we denote its support by
$\mbox{supp}\{q\}$, that is,
$\mbox{supp}\{q\}=\{\bx:~q(\bx)>0\}$.

For two positive sequences, $\{a_n\}_{n\ge 1}$ and $\{b_n\}_{n\ge 1}$, the notation $a_n\exe b_n$ will
stand for equality in the exponential scale, that is,
$\lim_{n\to\infty}\frac{1}{n}\log \frac{a_n}{b_n}=0$. Similarly,
$a_n\lexe b_n$ means that
$\limsup_{n\to\infty}\frac{1}{n}\log \frac{a_n}{b_n}\le 0$, and so on.
The indicator function
of an event $\calA$ will be denoted by $\calI\{A\}$. The notation $[x]_+$
will stand for $\max\{0,x\}$. Logarithms will be understood to be taken to the
natural base, $\mbox{e}$, unless specified otherwise. The Q-function is
defined as
\begin{equation}
Q(u)=\frac{1}{\sqrt{2\pi}}\int_u^\infty e^{-x^2/2}\mbox{d}x.
\end{equation}

\subsection{Setup}
\label{setup}

Consider the memoryless Gaussian channel,
\begin{equation}
Y_t=X_t+Z_t,~~~~t=1,2,\ldots
\end{equation}
where $\{X_t\}$ is a real-valued channel input signal, 
$\{Y_t\}$ is the channel output signal, and
where $\{Z_t\}$ is an i.i.d., zero-mean Gaussian noise process with variance $\sigma^2$.

Given a positive integer $n$, let us define a set $\calS_n\subset\reals^n$ of
allowable channel input vectors to be:
\begin{equation}
\label{constraints}
\calS_n=\left\{\bx:~\sum_{t=1}^n\phi_j(x_t)\le
n\Gamma_j,~j=1,\ldots,k\right\},
\end{equation}
where $k$ is a positive integer, $\phi_j(\cdot)$ are certain cost constraint functions and
$\Gamma_j$ are given constants, $j=1,\ldots,k$. Clearly, equality constraints
can be formally incorporated by defining pairs of inequality constraints that differ by
their signs, i.e.,
\begin{eqnarray}
\label{equality1}
\sum_{t=1}^n\phi_j(x_t)&\le& n\Gamma_j\\ 
\label{equality2}
\sum_{t=1}^n[-\phi_j(x_t)]&\le& n\cdot[-\Gamma_j].
\end{eqnarray}
In some of our derivations and
results we will allow
more general cost constraint functions, where each $\phi_j(\cdot)$ (or at least one of them) operates on a
sliding-window of $m$ channel input symbols ($m$ - positive integer) rather than on a single symbol.
In this case, the constraints that define $\calS_n$ would be of form
\begin{equation}
\sum_{t=m}^n\phi_j(x_{t-m+1}^t)\le n\Gamma_j,~~~~~j=1,2,\ldots,k.
\end{equation}
For the case $m=1$, we refer to the constraints that define $\calS_n$ as {\em
memoryless constraints}, whereas the case $m\ge 2$, will be referred to as the
case of {\em constraints with memory} or as {\em sliding-window constraints}.

The most common example of a memoryless constraint is the average power
constraint, where $k=1$ and $\phi_1(x)=x^2$ and $\Gamma_1=P$, the allowed
power. Another important example is the
case where, in addition to the average power constraint, a peak-power
constraint is imposed, i.e., $|x_t|\le A$ (for a given positive real $A$) for
all $t=1,2,\ldots,n$. In this case, $k=2$, $\phi_1$ is as above, and the peak-power
constraint can be accommodated within our framework by 
using the infinite square well (ISW) function,
\begin{equation}
\label{well}
\phi_2(x)=w(x)\dfn\left\{\begin{array}{ll}
0 & |x|\le A\\
\infty & |x|>A\end{array}\right.
\end{equation}
and $\Gamma_2=0$.
Useful examples of cost constraint functions with memory are those 
associated with 
correlation constraints, e.g., 
\begin{equation}
\label{correlationconstraint}
\sum_{t=\ell+1}^n x_tx_{t-\ell}\le n R_\ell,
\end{equation}
where $\ell$ is a fixed positive integer and $R_\ell$ is a given constant.
Other useful examples could be associated with limitations on the relative
frequency of
sign changes along the vector $\bx$, i.e.,
\begin{equation}
\label{frequencyofsignchangesconstraint}
\sum_{t=2}^n \calI\{\mbox{sgn}(x_tx_{t-1})=-1\}\le n\alpha,
\end{equation}
or, for example, a peak-power limitation on a filtered version of $\bx$, i.e.,
\begin{equation}
\label{iswafterfilterconstraint}
\sum_{t=m}^n w\left(\sum_{i=0}^{m-1}h_ix_{t-i}\right)\le 0,
\end{equation}
where $\{h_i\}_{i=0}^{m-1}$ is the filter's impulse response.
For binary input channels, sliding-window constraints can also be used to
limit (from above and/or below) the number of successive repetitions (or
run-lengths) of
certain channel input symbols (see, e.g., \cite{SK90}). To this end, one may
define the corresponding cost constraint function to be equal to zero for
every allowable channel input pattern and to be equal to infinity for every forbidden pattern.

Let us denote $\bGamma=(\Gamma_1,\ldots,\Gamma_k)$, and define the channel capacity
subject to the given constraints as
\begin{equation}
C(\bGamma)\dfn\liminf_{n\to\infty}\sup \frac{I(\bX;\bY)}{n},
\end{equation}
where the supremum is taken over all input distributions, $\{q\}$, with
$\mbox{supp}\{q\}\subseteq
\calS_n$. 

\subsection{Objective}
\label{objective}

The objective of this work is to propose two general
methodologies for obtaining fairly tight lower bounds to $C(\bGamma)$. 

The first methodology is based on the entropy power inequality (EPI)
and is therefore applicable only when the channel input vector, $\bX$,
takes on continuous values within $\calS_n$.
Since the EPI lower bound to mutual information depends on the input
distribution only via its differential entropy, $h(\bX)$, it is obvious that the
maximizing distribution for the EPI lower bound is uniform across $\calS_n$,
namely, $q(\bx)=1/\mbox{Vol}\{\calS_n\}$ for $\bx\in\calS_n$ and
$q(\bx)=0$ elsewhere. Consequently, the corresponding lower bound to
$C(\bGamma)$ hinges upon our ability to assess the volume of $\calS_n$, or
more precisely, the asymptotic exponential rate of $\mbox{Vol}\{\calS_n\}$ as
a function of $n$, namely, the {\em volume exponent} of $\{\calS_n,~n\ge 1\}$, 
defined as
\begin{equation}
\label{volumeexponentdef}
v(\bGamma)\dfn\lim_{n\to\infty}\frac{\log\mbox{Vol}\{\calS_n\}}{n} 
\end{equation}
for the general form
of $\calS_n$, defined by either memoryless constraints or sliding-window
constraints. Note that the limit of (\ref{volumeexponentdef}) exists due to
super-additivity of the sequence $\{\log\mbox{Vol}\{\calS_n\}\}_{n\ge 1}$.
To this end, we
invoke tools associated with the extended method of types and saddle-point integration
\cite[Chapters 2 and 3]{MW25} (see also \cite{deBruijn81}) as well as
elementary results from large
deviations theory \cite{DZ98}. 

The second methodology is based on direct
manipulation of the mutual information, $I(\bX;\bY)$, where instead of
maximizing over all input distributions supported by $\calS_n$, we take the
input distribution to be uniform across $\calS_n$, and once again, the
resulting lower bounds will depend on the volume exponent of $\calS_n$. This
class of bounds is somewhat more complicated than the EPI bound, but still reasonably simple and
easy to calculate at least for memoryless constraints. More importantly, it is
tighter and stronger than the EPI bound, as will be demonstrated in numerical
examples. It is also applicable for both discrete and continuous channel
inputs, unlike the EPI bound which is valid only for continuous inputs.
Moreover, it is easy to extend and strengthen this class of bounds
by allowing optimization over certain (parametric) classes of non-uniform input 
distributions across $\calS_n$. The
caveat, however, is that in order 
to handle constraints with memory, there is a need to further give
up on tightness at a certain point, for reasons that will become apparent in
the sequel.

\section{EPI Volume-Based Bounds}
\label{epi}

\subsection{Elementary Background on EPI Lower Bounds}
\label{basicepi}

The idea of deriving lower bounds to the channel capacity by invoking the EPI
is simple and not quite new, see, e.g., \cite{JA14}, \cite{JA16}, \cite{PS24}, \cite{SBD95},
\cite{TKB17}. Nonetheless, for the sake of completeness, we begin this section
by presenting it, and then combine it with our method \cite{MW25} for evaluating the
volume exponent, $v(\bGamma)$, of $\{\calS_n\}$.

For every channel input distribution, $q(\cdot)$, whose support is within
$\calS_n$, consider the following chain of inequalities:
\begin{eqnarray}
\frac{I(\bX;\bY)}{n}&=&\frac{h(\bY)-h(\bY|\bX)}{n}\nonumber\\
&=&\frac{h(\bY)}{n}-\frac{h(\bZ)}{n}\nonumber\\
&\ge&\frac{1}{2}\log\left[\mbox{e}^{2h(\bX)/n}+\mbox{e}^{2h(\bZ)/n}\right]-\frac{h(\bZ)}{n}\nonumber\\
&=&\frac{1}{2}\log\left[\mbox{e}^{2h(\bX)/n}+
2\pi\mbox{e}\sigma^2\right]-\frac{1}{2}\log(2\pi\mbox{e}\sigma^2)\nonumber\\
&=&\frac{1}{2}\log\left[1+\frac{\mbox{e}^{2h(\bX)/n}}{2\pi\mbox{e}\sigma^2}\right],
\end{eqnarray}
and so,
\begin{eqnarray}
C_n(\bGamma)&\dfn&\sup_{\{q:~\mbox{supp}\{q\}\subseteq\calS_n\}}\frac{I(\bX;\bY)}{n}\nonumber\\
&\ge&\sup_{\{q:~\mbox{supp}\{q\}\subseteq\calS_n\}}
\frac{1}{2}\log\left[1+\frac{\mbox{e}^{2h(\bX)/n}}{2\pi\mbox{e}\sigma^2}\right]\nonumber\\
&=&\frac{1}{2}\log\left[1+\frac{\exp\{2\log\mbox{Vol}\{\calS_n\}/n\}}{2\pi\mbox{e}\sigma^2}\right].
\end{eqnarray}
Taking the limit inferior of $n\to\infty$, we arrive at
\begin{equation}
C(\bGamma)\ge C_{\mbox{\tiny EPI}}(\bGamma)\dfn 
\frac{1}{2}\log\left[1+\frac{\exp\{2v(\bGamma)\}}{2\pi\mbox{e}\sigma^2}\right].
\end{equation}

\subsection{Volume Exponents}
\label{volume-exponent}

As mentioned at the very beginning of this section, this generic EPI lower
bound is known for some time.
For example, if only an average power constraint is imposed (i.e., $k=1$,
$\phi_1(x)=x^2$, and $\Gamma_1=P$), then $\calS_n$ 
is an $n$-dimensional Euclidean ball of radius $\sqrt{nP}$,
whose volume is given by
$\mbox{Vol}\{\calS_n\}=\pi^{n/2}(\sqrt{nP})^n/\Gamma(n/2+1)\exe (2\pi e
P)^{n/2}$ (with $\Gamma(\cdot)$ being the Gamma function, not to be confused
with the vector $\bGamma$ or its components), and so, $v(P)=\frac{1}{2}\log(2\pi e P)$, leading to the tight
lower bound,
\begin{eqnarray}
C_{\mbox{\tiny EPI}}(P)&=&\frac{1}{2}\log\left(1+\frac{e^{2v(P)}}{2\pi
e\sigma^2}\right)\nonumber\\
&=&\frac{1}{2}\log\left(1+\frac{2\pi e P}{2\pi e\sigma^2}\right)\nonumber\\
&=&\frac{1}{2}\log\left(1+\frac{P}{\sigma^2}\right)\nonumber\\
&=&C(P).
\end{eqnarray}
However, for the general case, where $\calS_n$ is defined by 
arbitrary sets of cost constraint functions, the evaluation of $v(\bGamma)$ seems to
be less trivial. This is exactly the point where our contribution in this section takes
place, as we invoke the techniques described in Chapters 2 and 3 of \cite{MW25}.

We begin from the case of memoryless constraints, and later discuss the
extension to constraints with memory.
Let $\btheta=(\theta_1,\ldots,\theta_k)$, where $\theta_i> 0$ for all
$i=1,2,\ldots,k$, and define the function
\begin{equation}
Z(\btheta)\dfn\int_{-\infty}^\infty\exp\left\{-\sum_{j=1}^k\theta_j\phi_j(x)\right\}\mbox{d}x,
\end{equation}
and assume that $Z(\btheta)<\infty$ for some subset $\Theta$ 
of $\reals_+^k$, which is the set of all $k$-vectors, $\btheta$, with strictly
positive components. In the discrete case, the integration
over $\reals$ is replaced by a summation over $\calX$, the alphabet of $x$.
For shorthand notation, in the sequel, we define the vector
function $\bphi(x)=(\phi_1(x),\ldots,\phi_k(x))$, and the inner product
$\btheta\bullet\bphi(x)=\sum_{j=1}^k\theta_j\phi_j(x)$, so that
we can write
$Z(\btheta)\dfn\int_{-\infty}^\infty\exp\{-\btheta\bullet\bphi(x)\}\mbox{d}x$.
We also denote the inner product
$\btheta\bullet\bGamma=\sum_{j=1}^k\theta_j\Gamma_j$.
Let us define
\begin{eqnarray}
\psi(\btheta)&\dfn&\log Z(\btheta),\\
\dot{\psi}(\btheta)&\dfn&\nabla\psi(\btheta)=\left(\frac{\partial\psi(\btheta)}{\partial\theta_1},\ldots,
\frac{\partial\psi(\btheta)}{\partial\theta_k}\right),\\
\ddot{\psi}(\btheta)&\dfn&\nabla^2\psi(\btheta),
\end{eqnarray}
where $\nabla^2\psi(\btheta)$ is the $k\times k$ Hessian matrix whose
$(i,j)$-th element is given by
$\frac{\partial^2\psi(\btheta)}{\partial\theta_i\partial\theta_j}$,
$i,j=1,\ldots,k$. Finally, define
\begin{equation}
\omega(\bGamma)\dfn\inf_{\btheta\in\reals_+^k}\{\btheta\bullet\bGamma+\psi(\btheta)\}.
\end{equation}

The following lemma, whose proof appears in the Appendix A, establishes the result that under certain regularity
conditions, the volume exponent,
$v(\bGamma)$, is equal to the function $\omega(\bGamma)$ defined above.

\begin{lemma}
\label{lemma1}
Let $\bphi(\cdot)$ be defined such that $Z(\btheta)<\infty$ for all
$\theta\in\Theta\in\reals_+^k$. Then,
\begin{enumerate}
\item $v(\bGamma)\le\omega(\bGamma)$.
\item Assume, in addition, that for the
given $\bGamma$, there
exists $\btheta$ such that $\dot{\psi}(\btheta)=-\bGamma$ and that
$\ddot{\psi}(\btheta)$ is a positive definite matrix with finite diagonal entries.
Then, 
\begin{equation}
v(\bGamma)\ge\omega(\bGamma), 
\end{equation}
and therefore, following part 1, 
\begin{equation}
v(\bGamma)=\omega(\bGamma). 
\end{equation}
\end{enumerate}
\end{lemma}

Since Lemma 1 plays a central role in our derivations, it is useful to pause 
and discuss both its significance and technical aspects before applying it to
obtain capacity bounds.\\

\noindent
1. {\em Maximum-entropy representation}. As discussed in \cite[pp.\ 40-41]{MW25}, 
it is not difficult to show that an equivalent expression for $v(\bGamma)$
is given by the maximum-entropy variational representation as the supermum of
the differential entropy, $h(X)$, of a random variable $X$ subject to the simultaneous 
constraints, $\bE\{\phi_j(X)\}\le\Gamma_j$, $j=1,2,\ldots,k$.\\

\noindent
2. {\em Equality constraints.} As mentioned earlier, an equality constraint can
be accommodated by a pair of inequality constraints with the same cost function and
cost limit, but with opposite signs (see eqs.\ (\ref{equality1}) and
(\ref{equality2}) above). This amounts to allowing the
corresponding parameter $\theta_j$ to take on any real value, not just
positive values, exactly as in constrained optimization using
Lagrangians.\\

\noindent
3. {\em Convex optimization.} It is easy to see that $\psi(\btheta)$ is a
convex function by observing that its Hessian, $\ddot{\psi}(\btheta)$, is non-negative definite
due to the fact that it can be viewed as a covariance matrix of the random
vector $\bphi(X)$ under $f_{\btheta}$, the probability density function (pdf) that is proportional to
$e^{-\btheta\bullet\bphi(x)}$. Therefore, the
minimization associated with the calculation of $v(\bGamma)$ is a convex
program, and hence can be calculated using standard tools of convex
programming.\\

\noindent
4. {\em Redundant constraints and their removal.} Part 2 of Lemma 1 assumes that we
can find $\btheta$ such that $\dot{\psi}(\btheta)=-\bGamma$. It might happen,
however, that
this condition is violated at certain instances of the problem. This may be
the case when either $\calS_n$ is empty (which is the case when the
constraints are contradictory), or when there are redundant
constraints, namely, inactive constraints, which are superfluous in the presence of
other constraints. Such constraints are characterized by holding with strict
inequalities, i.e., $\sum_{i=1}^n\phi_j(x_i)<\Gamma_i$.
As an example of a redundant constraint, consider the case $k=2$,
$\phi_1(x)=|x|$, and $\phi_2(x)=x^2$. Since
$\left(\frac{1}{n}\sum_{i=1}^n|x_i|\right)^2$ cannot exceed
$\frac{1}{n}\sum_{i=1}^nx_i^2$, it is obvious that the constraint
$\frac{1}{n}\sum_{i=1}^n|x_i|\le\Gamma_1$ becomes redundant whenever
$\Gamma_1>\sqrt{\Gamma_2}$. In general, a redundant constraint, 
$\sum_{i=1}^n\phi_j(x_i)<\bGamma_j$,
can be formally removed simply
by the assigning $\theta_j=0$.\\

\noindent
5. {\em Alternative methods.} In the appendix, we prove Lemma \ref{lemma1} by using standard
probabilistic arguments. One alternative technique involves the saddle-point
integration method (see \cite[Chapter 3]{MW25} and references therein).
In this approach, one first expresses the volume of $\calS_n$ as
\begin{equation}
\label{volumedef}
\mbox{Vol}\{\calS_n\}=\int_{\reals^n}\mbox{d}\bx\prod_{\ell=1}^kU\left(n\Gamma_\ell-\sum_{i=1}^n\phi_\ell(x_i)\right),
\end{equation}
where $U(t)$ is the unit step function. Then, one represents each factor,
$U\left(n\Gamma_\ell-\sum_{i=1}^n\phi_\ell(x_i)\right)$, of the integrand as the
inverse Laplace transform of $1/s$, computed at the point
$n\Gamma_j-\sum_{i=1}^n\phi_j(x_i)$, i.e.,
\begin{equation}
\label{inverselaplace}
U\left(n\Gamma_\ell-\sum_{i=1}^n\phi_\ell(x_i)\right)=\lim_{T\to\infty}\frac{1}{2\pi
j}\int_{c-jT}^{c+jT}\frac{\mbox{d}s}{s}\exp\left\{s\left(n\Gamma_\ell-\sum_{i=1}^n\phi_\ell(x_i)\right)\right\},
\end{equation}
where $j\dfn\sqrt{-1}$ and $c$ is any positive real. Finally, after substituting
(\ref{inverselaplace}) into (\ref{volumedef})
and interchanging the order of
integrations, one applies the saddle-point approximation to the resulting
integral in the (multivariate) complex plane (see Section 3.4 of \cite{MW25}). 

Another technique, that is sometimes applicable, is inspired by large deviations theory 
(see, e.g., \cite{DZ98}). Suppose, for example, that one of the constraints that define
$\calS_n$ involves the ISW function $w(\cdot)$ defined in (\ref{well}), which
means that $\calS_n\subseteq [-A,A]^n$, as will be the case in many of our
examples in the sequel. Then, one may imagine an auxiliary random
vector $\bX=(X_1,\ldots,X_n)$, uniformly distributed within $[-A,A]^n$, and then assess
the probability, $\mbox{Pr}\{\bX\in\calS_n\}\equiv\mbox{Vol}\{\calS_n\}/(2A)^n$, using the
Chernoff bound, which is exponentially tight under rather general conditions.
Then, the estimated volume of $\calS_n$ would be $(2A)^n$ times the
estimated probability of $\calS_n$, and so, $v(\bGamma)=\log(2A)-\bI(\bGamma)$,
$\bI(\bGamma)$ being the large-deviations rate function of the event $\calS_n$.\\

\noindent
6. {\em Analogy with statistical physics.} Readers familiar with elements of
statistical physics (others may skip this comment without loss of
continuity), may recognize the resemblance between the formula,
\begin{equation}
\label{volumeexponentformula}
v(\bGamma)=\inf_{\btheta\in\reals_+^k}\{\btheta\bullet\bGamma+\psi(\btheta)\}
\end{equation}
and the Legendre-Fenchel relationship between the
so called specific entropy of the micro-canonical ensemble, associated with
a system of $n$ particles subjected to the constraints that define $\calS_n$, and
the corresponding specific free energy, $-\psi(\btheta)$, of the equivalent canonical
(or Gibbs) ensemble, whose partition function is $Z(\btheta)$. 
Each component, $\theta_j$, of the parameter vector $\btheta$ has the physical
significance of a certain external
force (one of them being the inverse temperature) that is conjugate to the
corresponding macroscopic quantity, $\sum_{i=1}^n\phi_j(x_i)$. This external force is applied to the canonical
physical system in order to control the expectation of 
$\sum_{i=1}^n\phi_j(x_i)$, so as to keep it in compliance with the constraints of
$\calS_n$ (with high probability for large $n$). For more details regarding these
relations, see the discussion in the last paragraph of
Section 2.2 (page 44) in \cite{statmechbook}.

\subsection{Some Examples}
\label{someexamples}

In this subsection, we consider a few simple examples.

\noindent
{\bf Example 1 - simultaneous average power and peak power constraints.}
Let $k=2$, 
$\phi_1(x)=x^2$, $\Gamma_1=P$, 
$\phi_2(x)=w(x)$, and $\Gamma_2=0$. Then,
\begin{equation}
Z(\btheta)=\int_{-\infty}^\infty \exp\{-\theta_1
x^2-\theta_2w(x)\}\mbox{d}x=\int_{-A}^{A}e^{-\theta_1
x^2}\mbox{d}x=\sqrt{\frac{\pi}{\theta_1}}\cdot[1-2Q(A\sqrt{2\theta_1})].
\end{equation}
and so, the volume exponent is
\begin{eqnarray}
v(P)&=&\inf_{\theta_1>0}\left\{\theta_1P+\frac{1}{2}\log\frac{\pi}{\theta_1}+\log[1-2Q(A\sqrt{2\theta_1})]\right\}\nonumber\\
&=&\inf_{s>0}\left\{\frac{s}{2}+\frac{1}{2}\log\left(\frac{2\pi
P}{s}\right)+\log\left[1-2Q\left(\sqrt{\frac{A^2s}{P}}\right)\right]\right\},
\end{eqnarray}
where in the second line we have changed the optimization variable according
to $s=2\theta_1P$, with the benefit that in the resulting expression of $v(P))$ 
the dependence upon $A^2/P$ appears more explicitly.
It follows that the corresponding EPI lower bound is given by
\begin{eqnarray}
C_{\mbox{\tiny EPI}}(P,A)&=&\frac{1}{2}\log\left[1+\frac{e^{2v(P)}}{2\pi
e\sigma^2}\right]\\
&=&\frac{1}{2}\log\left[1+\frac{P}{\sigma^2}\cdot\inf_{s>0}\left\{\frac{e^{s-1}
}{s}\cdot\left[1-2Q\left(\sqrt{\frac{A^2s}{P}}\right)\right]^2\right\}\right]\nonumber\\
&\dfn&\frac{1}{2}\log\left[1+\frac{P}{\sigma^2}\cdot\lambda\left(\frac{A^2}{P}\right)\right].
\end{eqnarray}
The factor 
\begin{equation}
\lambda\left(\frac{A^2}{P}\right)=\inf_{s>0}\left\{\frac{e^{s-1}
}{s}\cdot\left[1-2Q\left(\sqrt{\frac{A^2s}{P}}\right)\right]^2\right\},
\end{equation}
which clearly depends only on the ratio $A^2/P$,
can be viewed as the effective factor of loss in signal-to-noise ratio (SNR) due to the peak-power
constraint, relative to the ordinary Gaussian
channel with average power constraint only. Indeed, it is easy to see that
$\lambda(u)$ never exceeds unity (by setting $s=1$ instead of minimizing over
$s$) and that $\lim_{u\to\infty}\lambda(u)=1$.

\begin{figure}[h!t!b!]
\centering
\includegraphics[width=8.5cm, height=8.5cm]{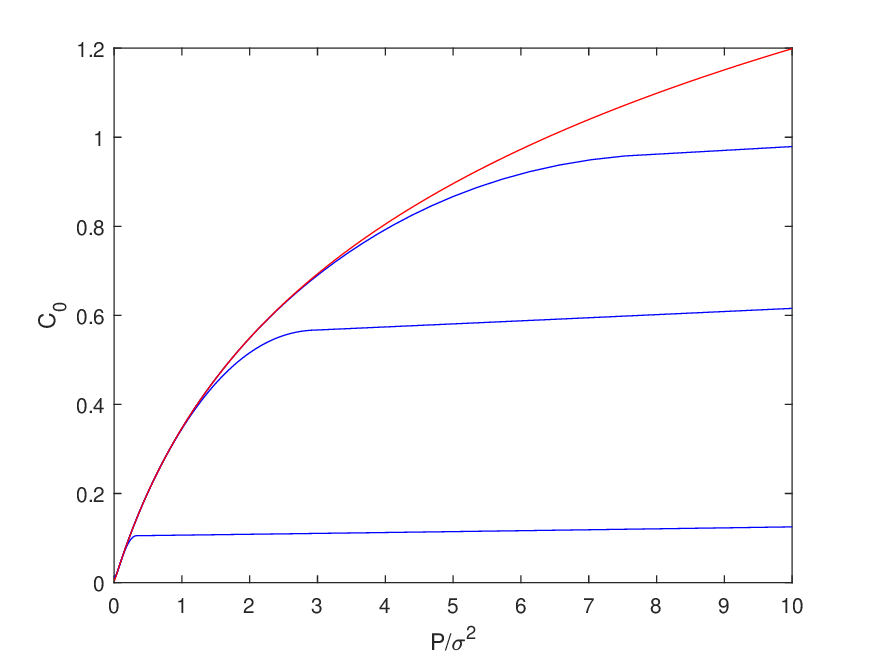}
\caption{$C_{\mbox{\tiny EPI}}(P,A)$ as a function of $P/\sigma^2$ (with $\sigma^2=1$) for $A=1$, $A=3$, $A=5$ (blue
curves) and $A=\infty$ (red curve).}
\label{graph1}
\end{figure}

\begin{figure}[h!t!b!]
\centering
\includegraphics[width=8.5cm, height=8.5cm]{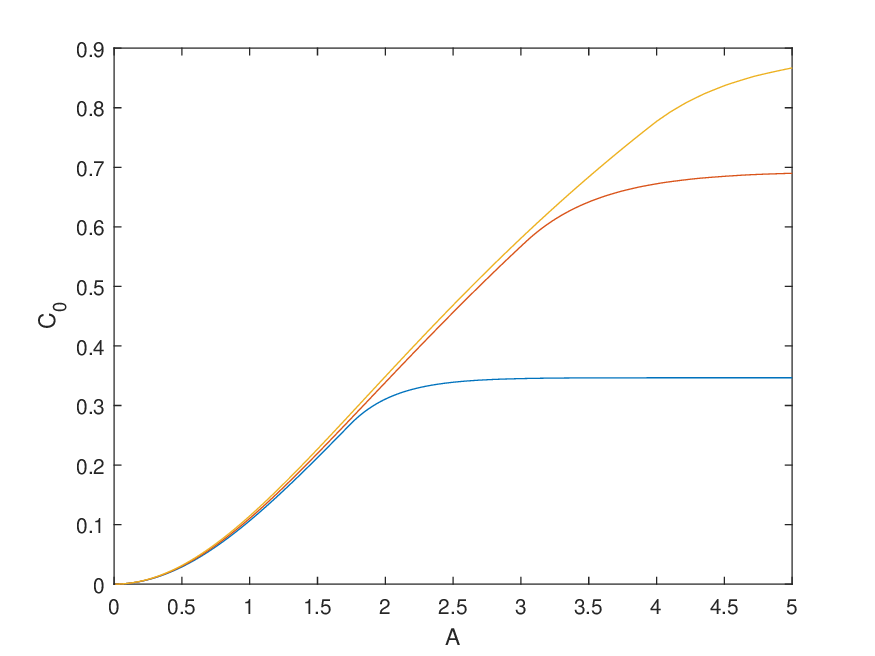}
\caption{$C_{\mbox{\tiny EPI}}(P,A)$ as a function of $A$ for $\sigma^2=1$, $P/\sigma^2=1$, $P/\sigma^2=3$, and
$P/\sigma^2=5$.}
\label{graph2}
\end{figure}

In Figures \ref{graph1} and \ref{graph2}, we display plots of $C_{\mbox{\tiny
EPI}}(P,A)$ as functions
of $P$ (in units of $\sigma^2$) for various values of $A$ and 
of $C_{\mbox{\tiny EPI}}(P,A)$
as functions of 
$A$ for various values of $P/\sigma^2$, respectively. 
It is interesting to compare the EPI bound with the exact capacity results due
to Smith \cite{Smith71} for this model. 
In Fig.\ 4 of \cite{Smith71}, Smith plots a curve of the exact capacity for
$A=\sqrt{2P}$ as a function of $\mbox{SNR}=P/\sigma^2$ in dB. Upon reading this
curve, one finds that
for $P/\sigma^2=10\mbox{dB}$, the real capacity is in the
vicinity of
1.1 nats/channel-use, whereas our EPI bound is 0.8688.
Likewise, for $P/\sigma^2=6\mbox{dB}$, the true capacity is approximately
0.802 nats/channel-use, while the EPI bound is 0.5262.
Finally, for $P/\sigma^2=12\mbox{dB}$, Smith's capacity is nearly
1.412 nats/channel-use, and the EPI bound is 1.0655. It should be stressed
that any loss of tightness in the EPI bound is solely due to the EPI, as the
volume exponent is exact. Later on we shall suggest several improved lower
bounds, which yield results much closer to the exact capacity of Smith's model.

An interesting two-dimensional version of this example concerns the quadrature channel. In this case,
suppose that $n$ is even, divide the components of $\bx$ into $n/2$ pairs
$(x_{2i-1},x_{2i})$, $i=1,2,\ldots,n/2$, and instead of the peak-power
constraint on each $x_i$, consider the
constraints $x_{2i-1}^2+x_{2i}^2\le A^2$ for all
$i=1,2\ldots,n/2$. The global
average power constraint $\sum_{i=1}^nx_i^2\le nP$ remains intact.
In this case, the partition function becomes
\begin{eqnarray}
Z(\theta)&=&\int_{\{(x_1,x_2):~x_1^2+x_2^2\le
A^2\}}e^{-\theta(x_1^2+x_2^2)}\mbox{d}x_1\mbox{d}x_2\nonumber\\
&=&\int_0^{2\pi}\int_0^A e^{-\theta r^2}r\mbox{d}r\mbox{d}\vartheta\nonumber\\
&=&2\pi\cdot\int_0^A e^{-\theta r^2}r\mbox{d}r\nonumber\\
&=&2\pi\cdot\int_0^{A^2/2} e^{-2\theta r^2/2}\mbox{d}\left(\frac{r^2}{2}\right)\nonumber\\
&=&2\pi\cdot\int_0^{A^2/2} e^{-2\theta u}\mbox{d}u\nonumber\\
&=&\frac{\pi}{\theta}\cdot(1-e^{-\theta A^2}).
\end{eqnarray}
The volume exponent is then
\begin{equation}
v(P)=\inf_{\theta\ge 0}\{\theta
P+\log\left(\frac{\pi}{\theta}\right)+\log(1-e^{-\theta A^2})\},
\end{equation}
and similarly as above,
\begin{equation}
C_{\mbox{\tiny EPI}}(P,A)=\frac{1}{2}\log\left[1+\frac{P}{\sigma^2}\cdot\inf_{s>0}\frac{
e^{s-1}}{s}\cdot\left(1-\exp\left\{-\frac{sA^2}{2P}\right\}\right)\right].
\end{equation}
Note that whenever $P\ge A^2/2$, the infimum is approached
for $s\to 0$, which yields
$C_{\mbox{\tiny EPI}}(P,A)=\frac{1}{2}\log(1+\frac{A^2}{2e\sigma^2})=
\frac{1}{2}\log(1+\frac{\pi A^2}{2\pi e\sigma^2})$, independent of $P$, as
expected, since the average power constraint becomes slack. The ``volume'' at
the numerator, $\pi
A^2$, is nothing but the area of a circle of radius $A$.  
This concludes Example 1. $\Box$\\

\noindent
{\bf Example 2 - absolute value constraint.}
In this example, we demonstrate that, in contrast to the true channel capacity function, $C(\bGamma)$, the lower
bound, $C_{\mbox{\tiny EPI}}(\bGamma)$, may not necessarily be a concave function. 
Consider the case $k=1$ with $\phi_1(x)=|x|$. In this case,
$Z(\theta)=2/\theta$, and so,
\begin{equation}
v(\Gamma)=\inf_{\theta>0}\left\{\theta\Gamma+\log\left(\frac{2}{\theta}\right)\right\}=\log(2e\Gamma),
\end{equation}
and so,
\begin{equation}
C_{\mbox{\tiny EPI}}(\Gamma)=
\frac{1}{2}\log\left(1+\frac{4e^2\Gamma^2}{2\pi e\sigma^2}\right)
=\frac{1}{2}\log\left(1+\frac{2e\Gamma^2}{\pi\sigma^2}\right)
=\frac{1}{2}\log\left(1+1.7305\frac{\Gamma^2}{\sigma^2}\right),
\end{equation}
which is obviously not concave, as for small $\Gamma$ it is nearly quadratic
(to the first order approximation):
\begin{equation}
C_{\mbox{\tiny EPI}}(\Gamma)\approx \frac{e}
{\pi\sigma^2}\cdot\Gamma^2,~~~~~~\Gamma\ll\sigma .
\end{equation}
More precisely, while 
the function $\log(1+\alpha x^2)$ $(\alpha> 0$ being a parameter) is concave in
$x>0$ across the range
$x\ge 1/\sqrt{\alpha}$, it is actually convex elsewhere.
Therefore the lower bound can be tightened by applying the upper concave
envelope (UCE) operator, namely,
\begin{equation}
\bar{C}_{\mbox{\tiny EPI}}(\bGamma)=
\mbox{UCE}\{C_{\mbox{\tiny EPI}}(\bGamma)\}\dfn\sup\left\{\sum_{i=1}^{k+1}\alpha_iC_{\mbox{\tiny EPI}}(\bGamma_i)\right\},
\end{equation}
where the supremum is over all assignments of
$(\bGamma_1,\ldots,\bGamma_{k+1})$ ($\bGamma_i$ being a $k$-vector for all
$i=1,2,\ldots,k+1$) and vectors
$\balpha=(\alpha_1,\ldots,\alpha_{k+1})$ with non-negative components such
that $\sum_{i=1}^{k+1}\alpha_i=1$ and
$\sum_{i=1}^{k+1}\alpha_i\bGamma_i=\bGamma$. The UCE,
$\bar{C}_{\mbox{\tiny EPI}}(\bGamma)$, can be achieved by time-sharing. In
this example, the UCE is obtained by replacing
$C_{\mbox{\tiny EPI}}(\Gamma)$ for small $\Gamma$ by a linear function, starting at the origin
and ending at the point $(\Gamma_\star,C_{\mbox{\tiny EPI}}(\Gamma_\star))$, where its
corresponding straight line is tangential to the curve of $C_{\mbox{\tiny EPI}}(\Gamma)$,
namely, the point pertaining to the non-zero solution to the equation $\Gamma
C_{\mbox{\tiny EPI}}'(\Gamma)=C_{\mbox{\tiny EPI}}(\Gamma)$, $C_{\mbox{\tiny
EPI}}'(\Gamma)$ being the derivative of
$C_{\mbox{\tiny EPI}}(\Gamma)$. The result is
\begin{equation}
\bar{C}_{\mbox{\tiny EPI}}(\Gamma)=\left\{\begin{array}{ll}
0.5293\cdot\frac{\Gamma}{\sigma} & \Gamma\le 1.5054\sigma\\
\frac{1}{2}\log\left(1+1.7305\cdot\frac{\Gamma^2}{\sigma^2}\right) & \Gamma\ge  1.5054\sigma\end{array}\right.
\end{equation}

\subsection{Constraints with Memory}
\label{slidingwindowconstraints}

The EPI lower bound, $C_{\mbox{\tiny EPI}}$ (as well as its UCE, if applicable), 
can also be extended to handle constraints with memory, or, sliding-window
constraints of the form (\ref{constraintswithmemory}), such as
(\ref{correlationconstraint}), (\ref{frequencyofsignchangesconstraint}),
(\ref{iswafterfilterconstraint}) as well as many others, where each
constraint function, $\phi_j$, manifests a certain limitation on the local
behavior of the channel input signal, for example, no more than $r$ sign changes
within each sliding window of length $m$ ($r<m$), or any other reasonable
criterion concerning the signal variability in the time domain.

First and foremost, we need an extended version of Lemma \ref{lemma1} to the
case where at least one of the constraint functions has memory $m\ge 2$.
Similarly as in Subsection \ref{volume-exponent}, it is useful to define a
parametric exponential family of densities, $f_{\btheta}(\cdot)$, except that
here these densities
would no longer be i.i.d., but densities derived from a Markov process of order
$m-1$, defined by
\begin{equation}
f_{\btheta}(\bx)=
\frac{\exp\left\{-\btheta\bullet\sum_{i=m}^n\bphi(x_{i-m+1}^i)\right\}}{Z_n(\btheta)}
=\frac{\prod_{i=m}^n\exp\left\{-\btheta\bullet\bphi(x_{i-m+1}^i)\right\}}{Z_n(\btheta)},
\end{equation}
where
\begin{equation}
Z_n(\btheta)\dfn\int_{\reals^n}\exp\left\{-\btheta\bullet\sum_{i=m}^n\bphi(x_{i-m+1}^i)\right\}\mbox{d}\bx.
\end{equation}
As before,
assume that $Z_n(\btheta)<\infty$ for every $\btheta\in\Theta\subseteq\reals_+^k$ and every
positive integer $n$. Assume further that the limit, $\lim_{n\to\infty}\frac{\log
Z_n(\btheta)}{n}$, exists and extend the definition of the function
$\psi(\btheta)$ to be
\begin{equation}
\psi(\btheta)\dfn\lim_{n\to\infty}\frac{\log Z_n(\btheta)}{n}.
\end{equation}
Now, part 1 of Lemma \ref{lemma1} extends straightforwardly under the new
definition of $\psi(\btheta)$. 
For part 2 of Lemma \ref{lemma1} to extend as well, 
we need to assume that: (i) $\psi(\btheta)$ is twice differentiable,
(ii) for the given $\bGamma$, there exists
$\btheta$ such that
$\dot{\psi}(\btheta)\dfn\nabla\psi(\btheta)=-(\bGamma-\epsilon)$,
for every sufficiently small $\epsilon>0$,
and (iii) the underlying Markov process corresponding to $f_{\btheta}$ is
(asymptotically) stationary and ergodic, so that by the ergodic theorem, for
every sufficiently small $\epsilon>0$,
\begin{equation}
\label{ergodicity}
\lim_{n\to\infty}\mbox{Pr}_{\btheta}\bigcap_{j=1}^k\left\{n(\Gamma_j-2\epsilon)\le\sum_{i=1}^n\phi_j(X_i)
\le n\Gamma_j\right\}=1,
\end{equation}
where $\mbox{Pr}_{\btheta}\{\cdot\}$ denotes probability under
$f_{\btheta}(\cdot)$, $\btheta$ being the point where
$\dot{\psi}(\btheta)=-(\bGamma-\epsilon)$.
In this case, the proof of the second part of Lemma \ref{lemma1} readily
generalizes to allow cost functions with memory, except that instead of using
the central limit theorem, we use (\ref{ergodicity}).

The main challenge here is to how calculate $\psi(\theta)$, which is required for
the calculation of the volume exponent, $v(\bGamma)$, according to
(\ref{volumeexponentformula}), but under the new definition of $\psi(\btheta)$.
Clearly, in the simple special case, where there are only autocorrelation
constraints, such as in (\ref{correlationconstraint}), without an additional
peak-amplitude constraint, the calculation of $Z_n(\btheta)$ involves a multivariate
Gaussian integral (with correlations) and
$C_{\mbox{\tiny EPI}}=\frac{1}{2}\log(1+\frac{\sigma_u^2}{\sigma^2})\}$, where $\sigma_u^2$ is the
innovation variance of an autoregressive process whose of order $m-1$ whose first $m$ autocorrelations are
$R_0,\ldots,R_{\ell-1}$ (see \cite{MW25}, Section 2.6). However, the general
case of cost functions with memory is more involved.

We now present two general methods for calculating $\psi(\btheta)$.\\

\noindent
1. {\em Integral operators and their spectral radius.}
According to this approach, we observe that
the calculation of $Z_n(\btheta)$ involves an assessment of the
exponential order of a multidimensional integral of the form:
\begin{equation}
I_n = \int_{-\infty}^\infty \cdots \int_{-\infty}^\infty \prod_{i=m}^n K(x_{i-m+1}^i)\mbox{d}\bx,
\end{equation}
where each factor of the integrand is the same kernel function
$K(x_{i-m+1}^i)=\exp\{-\btheta\bullet\bphi(x_{i-m+1}^i)\}$ applied to
a sliding window of 
$m$ variables, and where
$m$ remains fixed as $n \to \infty$. 
The calculation of $I_n$ can be viewed as a succession of iterated applications
of a sliding-window integral operator of the form
\begin{equation}
(\mathcal{L}g)(x_2^m) = 
\int_{-\infty}^\infty K(x^m)\cdot g(x^{m-1})\mbox{d}x_1.
\end{equation}
Accordingly, under mild regularity conditions, one may invoke the Collatz-Wielandt
formulas \cite{Collatz42}, \cite{Wielandt50} in order 
to assess the spectral radius of the operator $\mathcal{L}$, which coincides with
$e^{\psi(\btheta)}$.
These formulas are:
\begin{equation}
e^{\psi(\btheta)}=\inf_g\sup_{x^{m-1}}\frac{(\mathcal{L}g)(x^{m-1})}{g(x^{m-1})}=
\sup_g\inf_{x^{m-1}}\frac{(\mathcal{L}g)(x^{m-1})}{g(x^{m-1})}.
\end{equation}
These expressions can be viewed as continuous-alphabet counterparts of 
similar formulas for the Perron-Frobenius eigenvalue for finite dimensional
positive matrices in the finite-alphabet case. The second formula is somewhat more appealing in the sense that for the
purpose of obtaining a lower bound, it is legitimate to select an arbitrary
function $g$ rather than taking the supremum. Moreover, we have the freedom to choose a
parametric family of functions, say $\{g_\alpha\}$, which are convenient to work
with (like multivariate Gaussian
functions), and maximize only w.r.t.\ the parameter $\alpha$.
A similar comment applies w.r.t.\ $x^{m-1}$ in the first formula, but
optimization over a finite dimensional vector is less problematic than
optimization over a function, which involves, in general, calculus of
variations.

It should be noted that in the case $m=2$, if $K(\cdot,\cdot)$ is a symmetric
kernel, that is, $K(x,x')=K(x',x)$ for all $x$ and $x'$, then an
alternative formula for the spectral radius is the Rayleigh quotient formula,
\begin{eqnarray}
e^{\psi(\btheta)}&=&\sup_{u(\cdot)}\frac{\int_{-\infty}^\infty\int_{-\infty}^\infty
u(x)K(x,x')u(x')\mbox{d}x\mbox{d}x'}{\int_{-\infty}^\infty u^2(x)\mbox{d}x}\\
&=&\sup_{\{u(\cdot):~\|u\|^2=1\}}\int_{-\infty}^\infty\int_{-\infty}^\infty
u(x)K(x,x')u(x')\mbox{d}x\mbox{d}x'.
\end{eqnarray}
Once again, for the purpose of obtaining a valid lower bound, it is not
mandatory to calculate the supremum, and it is legitimate to pick an arbitrary
$u(\cdot)$ or to maximize within a parametric family, $\{u_\alpha(\cdot)\}$.\\

\noindent
{\em 2. The Donsker-Varadhan variational formula.}
Another possible characterization of the spectral radius, $e^{\psi(\btheta)}$, is via the
Donsker-Varadhan formula. Consider the Markov process of order $m-1$, whose
transition density is given by
\begin{equation}
f_{\btheta}(x_m|x^{m-1})=\frac{K(x^m)}{\int_{-\infty}^\infty
K(x^m)\mbox{d}x_m}\dfn
\frac{K(x^m)}{\exp\{S(x^{m-1})\}}.
\end{equation}
Then, neglecting edge effects for $n\gg m$,
\begin{eqnarray}
Z_n(\theta)&=&\int_{\reals^n}\prod_{i=m}^n
K(x_{i-m+1}^i)\mbox{d}\bx\nonumber\\
&=&\int_{\reals^n}\prod_{i=m}^nf_{\btheta}(x_i|x_{i-m+1}^{i-1})\exp\{
S(x_{i-m+1}^{i-1})\}\mbox{d}\bx\nonumber\\
&=&\bE_{\btheta}\left\{\exp\left[\sum_{i=m}^nS(X_{i-m+1}^{i-1})\right]\right\}\nonumber\\
&\eqa&\exp\left\{n\sup_g\bE_g\left[S(X^{m-1})-D(g(\cdot|X^{m-1})\|f_{\btheta}(\cdot|X^{m-1}))\right]\right\}\nonumber\\
&\eqb&\exp\left[n\sup_g\left(\bE_g\{\log
K(X^m)\}+h_g(X_m|X^{m-1})\right)\right]\nonumber\\
&=&\exp\left\{n\sup_g\left[-\btheta\bullet\bE_g\{\bphi(X^m)\}+h_g(X_m|X^{m-1})\right]\right\},
\end{eqnarray}
where (a) is due to the Donsker-Varadhan variational formula \cite{DV83}, 
(b) is by the definition of $S(\cdot)$, and
$h_g(X_m|X_{m-1})$ is the conditional differential entropy of $X_m$ given
$X^{m-1}$ under $g$. Consequently,
\begin{equation}
\psi(\btheta)=
\sup_g\left[-\btheta\bullet\bE_g\{\bphi(X^m)\}+h_q(X_m|X^m)\right],
\end{equation}
where for a given auxiliary $(m-1)$-th order Markov process $g$, the expectation,
$\bE_g\{\cdot\}$, is w.r.t.\ the stationary state
distribution associated with $g$.
Once again, for the purpose of obtaining a lower bound to
$\psi(\btheta)$, one may pick a particular $g$ or maximize within a
parametric family, $\{g_\alpha\}$ to facilitate the calculation at the
possible expense of losing tightness. For example, consider the conditional
pdf,
\begin{equation}
g^\star(x_m|x^{m-1})=\frac{\exp\left\{-\btheta\bullet\bphi(x^m)\right\}}{Z(\btheta,x^{m-1})},
\end{equation}
where
\begin{equation}
Z(\btheta,x^{m-1})=\int_{-\infty}^\infty\exp\left\{-\btheta\bullet\bphi(x^m)\right\}\mbox{d}x_m.
\end{equation}
Then,
\begin{equation}
\psi(\btheta)\ge\bE\log Z(\btheta,X^{m-1}),
\end{equation}
where the expectation is w.r.t.\ the stationary distribution associated with
the Markov process defined by $g^\star$. The reason that this is just a lower
bound is that $g^\star$ may not be the maximizer as its derivation does not take
into account the complicated dependence of the stationary distribution upon
the conditional densities of the underlying Markov process.\\

\subsection{More Examples}
\label{moreexamples}

We now provide a few examples for the case of cost functions with memory.\\

\noindent
{\bf Example 3 -- Peak-power limitation combined with a correlation constraint.} 
Consider the case of peak-power limitation combined with a
specified one-lag empirical autocorrelation $\sum_{t=2}^nx_tx_{t-1}=nR_1$. In this
case, $m=2$ and the partition function pertaining to the volume exponent is given by
\begin{eqnarray}
Z_n(\theta)&=&\int_{\reals^n}\exp[-\phi_1(x_1)]\prod_{t=2}^n\exp\{-w(x_t)-\theta
x_tx_{t-1}\}\mbox{d}\bx\nonumber\\
&=&\int_{[-A,A]^n}\prod_{t=2}^ne^{-\theta
x_tx_{t-1}}\mbox{d}\bx,
\end{eqnarray}
The exponential growth rate of $Z_n(\theta)$ is according to the largest
eigenvalue of the kernel $K(x,x')=e^{-\theta xx'}$ defined on the square
$(x,x')\in[-A,A]^2$.
Since the kernel $K(x,x')$ is symmetric,
the largest eigenvalue can be calculated using the Rayleigh quotient formula,
\begin{equation}
e^{\psi(\btheta)}= \sup_{\{u:~\|u\|^2=1\}} \int_{-A}^{A} \int_{-A}^{A}
u(x)e^{-\theta xx'}u(x')
\mbox{d}x\mbox{d}x'.
\end{equation}
To estimate $e^{\psi(\btheta)}$, we evaluate the Rayleigh quotient for the constant test function
$u(x) = \frac{1}{\sqrt{2A}}$, $x\in[-A,A]$,
which is normalized in $L^2([-A,A])$.
Thus,
\begin{equation}
e^{\psi(\btheta)}\geq \int_{-A}^{A} \int_{-A}^{A} \frac{1}{\sqrt{2A}} e^{-\theta
xx'} \frac{1}{\sqrt{2A}} \mbox{d}x\mbox{d}x'
= \frac{1}{2A} \int_{-A}^{A} \int_{-A}^{A} e^{-\theta x x'} \mbox{d}x
\mbox{d}x'.
\end{equation}
To evaluate the double integral
\begin{equation}
I = \int_{-A}^{A} \int_{-A}^{A} e^{-\theta x x'} \mbox{d}x \mbox{d}x',
\end{equation}
we can simplify it to a single integral as
\begin{equation}
\int_{-A}^{A} \int_{-A}^{A} e^{-\theta x x'} \mbox{d}x\mbox{d}x' =
\frac{4}{|\theta|} \int_0^{A^2|\theta|} \frac{\sinh u}{u} \mbox{d}u, 
\end{equation}
as can be shown by carrying out explicitly one of the two integrations of the
exponential function, and then changing the other integration variable.
Substituting into our earlier inequality, we find:
\begin{equation}
e^{\psi(\theta)}\geq \frac{1}{2A} \cdot \frac{4}{|\theta|} \int_0^{A^2|\theta|}
\frac{\sinh u}{u} \mbox{d}u 
= \frac{2}{A|\theta|} \int_0^{A^2|\theta|} \frac{\sinh u}{u} \mbox{d}u.
\end{equation}
We conclude that
\begin{equation}
e^{\psi(\btheta)}\ge \frac{2}{A|\theta|} \cdot
\operatorname{Shi}(A^2|\theta|),
\end{equation}
where $\operatorname{Shi}(z) = \int_0^z \frac{\sinh u}{u} \mbox{d}u$ 
is the hyperbolic sine integral function. This approximation is expected to be tight 
due to the positivity and symmetry of the kernel, and the choice of the constant function 
as a good candidate for the leading eigenfunction.
The resulting lower bound to the capacity is then given by
\begin{equation}
C_{\mbox{\tiny EPI}}=\frac{1}{2}\log\left[1+
\frac{\inf_{\theta\in\reals}e^{2R_1\theta}\frac{2}{A|\theta|} \cdot
\operatorname{Shi}(A^2|\theta|)}{2\pi
e\sigma^2}\right].
\end{equation}
More generally, we may bound the spectral radius of $K$ from below by considering the
parametric family
\begin{equation}
u_\alpha(x)=\sqrt{\frac{\alpha}{\sinh(2\alpha A)}}\cdot e^{-\alpha
x},~~~~~~|x|\le A,
\end{equation}
and then
\begin{equation}
e^{\psi(\theta)}\ge \sup_{\alpha\in\reals} \frac{2\alpha}{\sinh(2\alpha
A)}\int_{-A}^{A}\frac{\sinh(\theta x+\alpha)}{\theta x+\alpha}\mbox{d}x.
\end{equation}
The above example can easily be extended to accommodate also an additional
average power constraint (thus extending Example 1). In this case, the kernel
becomes $K(x,\hat{x})=\exp\{-\theta_1(x^2+\hat{x}^2)/2-\theta_2x\hat{x}\}$, which is still
symmetric.
This concludes Example 3. $\Box$\\

\noindent
{\bf Example 4 - Average power constraint combined with a correlation
constraint,} Consider the case $m=k=2$ with
$\phi_1(x_1,x_2)=\frac{x_1^2+x_2^2}{2}$, $\Gamma_1=P$,
$\phi_2(x_1,x_2)=x_1x_2$, and $\Gamma_2=\rho P$, where $|\rho|<1$.
In this case, denoting $s=\theta_2/\theta_1$, we have
\begin{equation}
Z(\btheta.x_1)=\int_{-\infty}^\infty
\exp\left\{-\theta_1(x_1^2+x_2^2)/2-\theta_2x_1x_2\right\}=\sqrt{\frac{2\pi}{\theta_1}}\cdot\exp\{-\theta_1(1-s^2)x_1^2/2\}.
\end{equation}
Taking the natural logarithm, then the expectation, and finally exponentiating
again, we obtain:
\begin{equation}
\exp\left[\bE\{\log
Z(\btheta,X_1)\}\right]\ge
\sqrt{\frac{2\pi}{\theta_1}}\cdot\exp\{-\theta_1(1-s^2)P/2\},
\end{equation}
where we have used the fact that $\bE_g\{X_1^2\}\le P$ (combined with the
conditions $\theta_1\ge
0$ and $s^2\le 1$), since $g$ must satisfy the moment constraints, as can be deduced from the
equivalent maximin problem of $\sup_g\inf_\theta\{\cdot\}$. Thus,
by standard optimization methods,
\begin{equation}
\lim_{n\to\infty}[\mbox{Vol}\{\calS_n\}]^{2/n}\ge \inf_{\theta_1>0,s}
\frac{2\pi}{\theta_1}\cdot\exp\{-\theta_1(1-s^2)P\}\cdot\exp\{2\theta_1P+2\theta_1s\rho
P\}=2\pi eP(1-\rho^2),
\end{equation}
and so,
$C_{\mbox{\tiny EPI}}\ge\frac{1}{2}\log\left[1+\frac{P(1-\rho^2)}{\sigma^2}\right]$, which is in
fact, the exact capacity under these constraints.$\Box$\\

\noindent
{\bf Example 5 - Peak amplitude limitation at the output of a linear system.}
Consider the case where the transmitter includes a linear filter just before
the antenna, and then the
peak-amplitude limitation applies to the filter output, namely, one the
constraints is (\ref{iswafterfilterconstraint}).
In this case the body defined by the constraints in the channel input space is
linearly transformed into an image body that resides in the filter output signal space, and
so the uniform distribution within the input body is transformed into a
uniform distribution across the image body. The volume of the image body is given by
the volume of the input body, multiplied by the Jacobian of the
transformation matrix. For a causal filter, this Jacobian is given by $|h_0|^n$.
Consequently, since the body at the filter output space is the hypercube 
$[-A,A]^n$, whose volume is $(2A)^n$, then the volume of its inverse image, at
the filter input space is $(2A)^n/|h_0|^n=(2A/|h_0|)^n$.
Note that, if this causal filter is also minimum phase,
then an alternative
expression for the Jacobian, in the frequency domain, is given by
\begin{equation}
\exp\left\{\frac{n}{2\pi}\int_{-\pi}^{\pi}\log|H(e^{j\omega})|\mbox{d}\omega\right\}.
\end{equation}

\section{Direct Manipulation of the Mutual Information}
\label{directmanipulation}

The EPI lower bounds are relatively simple and easy to calculate, provided that the
calculation of the volume exponent is reasonably easy. Lemma \ref{lemma1}
proposes a simple formula for the volume exponent, $v(\bGamma)$, which once
computed, it can be simply
substituted into the expression $\frac{1}{2}\log\left(1+\frac{e^{2v(\bGamma)}}{2\pi
e\sigma^2}\right)$ and the tightness of this bound depends solely on the
tightness of the EPI. However, the EPI is not always tight, especially not in
the range of low and moderate SNR. Furthermore, one of the severe limitations
of the EPI is that it applies merely to continuous-valued input vectors, and
not to discrete ones. 

In this section, we derive alternative families of lower
bounds, which are not based on the EPI, but rather on direct manipulation of
the mutual information, $I(\bX;\bY)$. The calculations of these bounds are somewhat more 
involved than that of the EPI bound, but it has the following advantages
compared to the EPI bound: (i) it is typically tighter, (ii) it applies to
both continuous and discrete channel inputs (with integrations over the
channel input space simply being replaced by summations), and (iii) it is easy to extend to
Gaussian channels with memory as well as to memoryless non-Gaussian channels.

\subsection{Memoryless Constraints}
\label{memorylessconstraints2}

As in Section \ref{epi},
we start from the case of memoryless cost functions.
Consider the following analysis of the mutual information, where instead of
maximizing the channel input pdf over the support $\calS_n$, we set the
uniform input pdf across $\calS_n$.
\begin{eqnarray}
I(\bX;\bY)&=&h(\bY)-\frac{n}{2}\log(2\pi e\sigma^2)\\
&=&-\bE\left\{\log\left[\frac{1}{\mbox{Vol}(\calS_n)}\int_{\calS_n}p(\bY|\bx)\mbox{d}\bx\right]\right\}-
\frac{n}{2}\log(2\pi e\sigma^2)\\
&=&-\bE\left\{\log\left[\int_{\calS_n}p(\bY|\bx)\mbox{d}\bx\right]\right\}+\log\mbox{Vol}(\calS_n)-\frac{n}{2}\log(2\pi
e\sigma^2).
\end{eqnarray}
Now, applying the Chernoff bounding technique, we have
\begin{eqnarray}
& &\bE\left\{\log\left[\int_{\calS_n}p(\bY|\bx)\mbox{d}\bx\right]\right\}\nonumber\\
&\le&\inf_{\btheta\in\reals_+^k}\bE\left\{\log\left[\int_{\reals^n}
\exp\left\{n[\btheta\bullet\bGamma-\btheta\bullet\sum_{i=1}^n\bphi(x_i)\right\}p(\bY|\bx)\mbox{d}\bx\right]\right\}\nonumber\\
&=&\inf_{\btheta\in\reals_+^k}\bE\left\{\log\left[\int_{\reals^n}
\exp\left\{n[\btheta\bullet\bGamma-\btheta\bullet\sum_{i=1}^n\bphi(x_i)\right\}
\frac{\exp\left\{-\|\bY-\bx\|^2/(2\sigma^2)\right\}}{(2\pi\sigma^2)^{n/2}}\mbox{d}\bx\right]\right\}\nonumber\\
&=&\inf_{\btheta\in\reals_+^k}\left[n\btheta\bullet\bGamma+\bE\left\{\log\left(\prod_{i=1}^n\left[\int_{\reals}
\frac{\exp\left\{-\btheta\bullet\bphi(x)
-(Y_i-x)^2/(2\sigma^2)\right\}}{\sqrt{2\pi\sigma^2}}\mbox{d}x\right]\right)\right\}\right]\nonumber\\
&\dfn&\inf_{\btheta\in\reals^k}\left[n\btheta\bullet\bGamma+
\bE\left\{\log\left(\prod_{i=1}^n\zeta_{\btheta}(Y_i)\right)\right\}\right]\nonumber\\
&=&n\cdot\inf_{\btheta\in\reals_+^k}\left(\btheta\bullet\bGamma+\bE\left\{\log
\zeta_{\btheta}(Y)\right\}\right), 
\end{eqnarray}
and so,
\begin{eqnarray}
I(\bX;\bY)&\ge& \log\mbox{Vol}(\calS_n)-\frac{n}{2}\log(2\pi
e\sigma^2)-n\cdot\inf_{\btheta\in\reals^K}\left(\btheta\bullet\bGamma+\bE\left\{\log
\zeta_{\btheta}(Y)\right\}\right)\nonumber\\
&=&\frac{n}{2}\log\frac{P}{\sigma^2}-
n\cdot\inf_{\btheta\in\reals^+}\left(\btheta\bullet\bGamma+\bE\left\{\log
\zeta_{\btheta}(Y)\right\}\right).
\end{eqnarray}
We have therefore arrived at the following 
lower bound:
\begin{equation}
\label{betterlowerbound}
C(\bGamma)\ge v(\bGamma)-
\frac{1}{2}\log(2\pi e\sigma^2)-
\inf_{\btheta\in\reals_+^k}\left(\btheta\bullet\bGamma+\bE\left\{\log
\zeta_{\btheta}(Y)\right\}\right),
\end{equation}
where
\begin{equation}
\zeta_{\btheta}(y)=\int_{\reals}
\frac{\exp\left\{-\btheta\bullet\bphi(x)
-(y-x)^2/(2\sigma^2)\right\}}{\sqrt{2\pi\sigma^2}}\mbox{d}x.
\end{equation}
Note that the second term of the lower bound to $C(\bGamma)$ requires
knowledge of the asymptotic marginal pdf of a single channel
output symbol, $Y$, 
which is given by the convolution between the pdf of a
single noise variable, namely, $\calN(0,\sigma^2)$, and the marginal of a single
component of the vector $\bX$, induced from the uniform pdf of $\bX$
across $\calS_n$ (for $n\to\infty$).
We will address this issue in the sequel through examples.\\

\noindent
{\bf Example 6 - Example 1 revisited.}
Consider again the case where there is both an average power constraint, $P$,
and a peak-power constraint,
$|x_i|\le A$, $i=1,2,\ldots,n$. As we showed in Example 1,
\begin{equation}
v(P)=\frac{1}{2}\log\left[2\pi
eP\cdot\inf_{s>0}\left\{\frac{e^{s-1}}{s}\left[1-2Q\left(s\sqrt{\frac{A^2}{P}}\right)\right]^2\right\}\right]
=\frac{1}{2}\log\left[2\pi
eP\cdot\lambda\left(\frac{A^2}{P}\right)\right].
\end{equation}
Now, after a straightforward algebraic manipulation, we obtain
\begin{eqnarray}
\log \zeta_\theta(y)&=&\log\left(\frac{1}{\sqrt{2\pi\sigma^2}}\int_{-A}^A\mbox{d}x\exp\left\{-\theta
x^2-\frac{(y-x)^2}{2\sigma^2}\right\}\right)\nonumber\\
&=&-\frac{1}{2}\log(1+2\theta\sigma^2)-\frac{\theta(\sigma_X^2+\sigma^2)}{1+2\theta\sigma^2}+\nonumber\\
& &\log\left[1-Q\left(\frac{A-m(y)}{s}\right)-
Q\left(\frac{A+m(y)}{s}\right)\right],
\end{eqnarray}
where
\begin{eqnarray}
s&=&\frac{\sigma}{\sqrt{1+2\theta\sigma^2}}\\
m(y)&=&\frac{y}{1+2\theta\sigma^2}\\
\sigma_X^2&=&\frac{\int_{-A}^Ax^2e^{-x^2/(2P)}\mbox{d}x}{\int_{-A}^Ae^{-x^2/(2P)}\mbox{d}x}.
\end{eqnarray}
Thus,
\begin{eqnarray}
C(\bGamma)&\ge&
\frac{1}{2}\log\left[\frac{P}{\sigma^2}\cdot\lambda\left(\frac{A^2}{P}\right)\right]-\nonumber\\
& &\inf_{\theta>0}\bigg\{\theta
P-\frac{1}{2}\log(1+2\theta\sigma^2)-\frac{\theta(\sigma_X^2+\sigma^2)}{1+2\theta\sigma^2}+\nonumber\\
& &\bE\log\left[1-Q\left(\frac{A-m(Y)}{s}\right)-Q\left(\frac{A+m(Y)}{s}\right)\right]\bigg\},
\end{eqnarray}
and the expectation is over the randomness of a single symbol, $Y$, given by
the convolution between the pdf of a single symbol $X$ and
$\calN(0,\sigma^2)$.
To determine the asymptotic marginal pdf of a single component, $X$, consider the following line
of thought (which is similar to the derivation of the Boltzmann distribution in statistical
mechanics): Given that $X=x$ ($|x|\le A$), the marginal pdf is proportional to the volume of
the body formed by the intersection between the hypercube $[-A,A]^{n-1}$ and
the hyper-ball $\calB_{n-1}(\sqrt{nP-x^2})$, which is given exponentially by
$$\exp\left\{\inf_{\theta\ge
0}\left[\theta(nP-x^2)+n\log\int_{-A}^{A}e^{-\theta\xi^2}\mbox{d}\xi\right]\right\},$$
which in turn, is proportional to $\exp\{-\theta_\star x^2\}$, where
$\theta_\star$ is the minimizer of $\theta
P+\log\int_{-A}^{A}e^{-\theta\xi^2}\mbox{d}\xi$. This minimizer is
$\theta_\star=0$ if $P\ge A^2/3$, and the unique positive solution to the equation
$$\frac{\int_A^Ax^2e^{-\theta x^2}\mbox{d}x}
{\int_A^A e^{-\theta x^2}\mbox{d}x}=P$$
if $P< A^2/3$.
Consequently, the asymptotic marginal of $X$ is uniform across $[-A,A]$
whenever $P\ge A^2/3$ and 
\begin{equation}
p_X(x)=\left\{\begin{array}{ll}
\frac{e^{-\theta_\star x^2}}{\sqrt{\pi/\theta_\star}[1-2Q(A\sqrt{2\theta_\star})]} & |x|\le A\\
0 & \mbox{elsewhere}\end{array}\right.
\end{equation}
whenever $P<A^2/3$.
Thus, defining $P'=\frac{1}{2\theta_\star}$, the asymptotic marginal of $Y$ is given by
\begin{equation}
p_Y(y)=\frac{\exp\left\{-\frac{y^2}{2(P'+\sigma^2)}\right\}}{\sqrt{2\pi(P'+\sigma^2)}[1-2Q(A/\sqrt{P'})]}\cdot
\left[Q\left(\frac{\zeta y-A}{\sigma_{\mbox{\tiny e}}}\right)-
Q\left(\frac{\eta y+A}{\sigma_{\mbox{\tiny e}}}\right)\right],
\end{equation}
with $\eta=\frac{P'}{P'+\sigma^2}$ and $\sigma_{\mbox{\tiny
e}}=\sqrt{\frac{P'\sigma^2}{P'+\sigma^2}}$.
In other words, the last term is given by
\begin{eqnarray}
& &\bE\log\left[1-Q\left(\frac{A-m(Y)}{s}\right)-Q\left(\frac{A+m(Y)}{s}\right)\right]\nonumber\\
&=&\int_{-\infty}^\infty\mbox{d}y P_Y(y)
\cdot\log\left[1-Q\left(\frac{A-m(y)}{s}\right)-Q\left(\frac{A+m(y)}{s}\right)\right].\nonumber
\end{eqnarray}

\begin{figure}[h!t!b!]
\centering
\includegraphics[width=8.5cm, height=8.5cm]{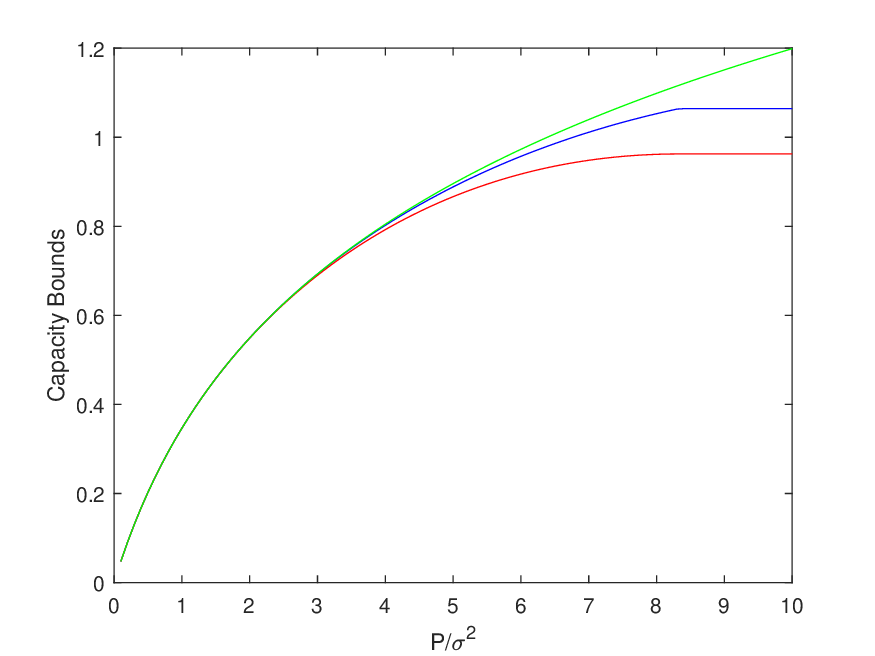}
\caption{Exact capacity of $A=\infty$ (green), EPI-based bound (red), and
present bound (blue), both for $A=5$ and $\sigma^2=1$.} 
\label{graph3}
\end{figure}

As can be seen in Fig.\ \ref{graph3}, the present bound is better than
the EPI bound, and the gap becomes visible especially as SNR grows. Note that
in the blue graph there is a phase transition at $P=A^2/3=5^2/3=8.333...$
beyond which the power constraint becomes slack given the amplitude
constraint.

Let us also revisit the comparison Smith \cite{Smith71}, but this time, also with
numerical results on the present bound. As mentioned earlier, Fig.\ 4 of his paper
displays plots of the capacity as function of the the SNR for
$A=\sqrt{2P}$. For $P/\sigma^2=10\mbox{dB}$, the exact capacity is approximately
1.1 nats/channel-use, the EPI bound is 0.8688 and the present bound
gives 0.9743. 
Likewise, for $P/\sigma^2=6\mbox{dB}$, the true capacity is nearly
0.802 nats/channel-use, the EPI bound is 0.5262 and the new bound
gives 0.6316. 
Finally, for $P/\sigma^2=12\mbox{dB}$, Smith's capacity is approximately
1.412 nats/channel-use, but the EPI bound is 1.0655 and the current bound
gives 1.1626. 

\begin{figure}[h!t!b!]
\centering
\includegraphics[width=8.5cm, height=8.5cm]{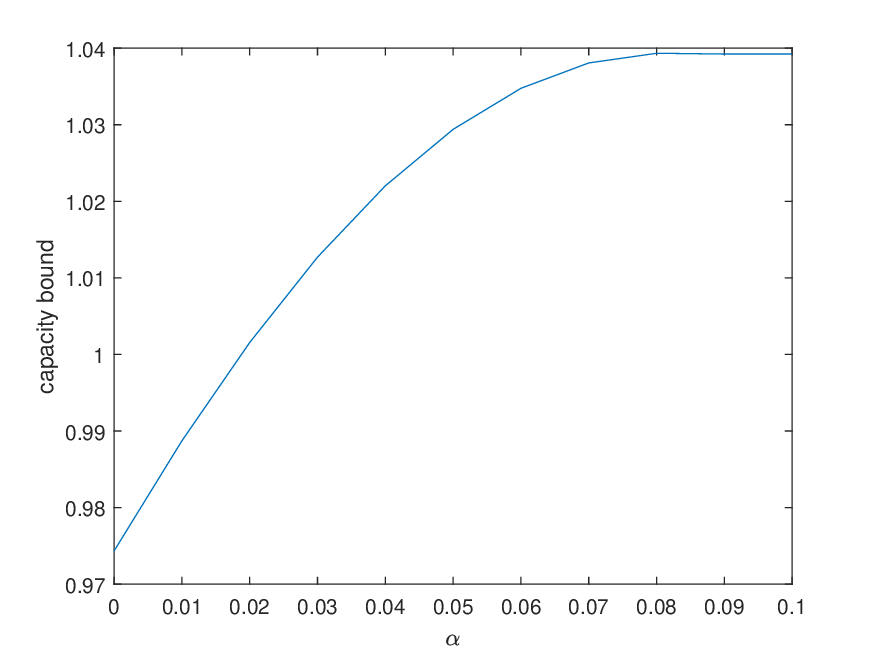}
\caption{Present bound for $A=\sqrt{20}$, $P=10$ and $\sigma^2=1$ as a
function of $\alpha$.}
\label{graph4}
\end{figure}

Returning to the case $P/\sigma^2=10\mbox{dB}$, we also extended
our present lower bound to correspond to 
$q(\bx)\propto
e^{\alpha\|\bx\|^2}$ within $\calS_n$ and zero elsewhere (see derivation in
Appendix B), and as expected, $\alpha>0$
improves on the uniform pdf within $\calS_n$ ($\alpha=0$), since higher energy
input vectors are preferred. This
improved our lower bound from 0.9743 of $\alpha=0$ up to 1.0393 for $\alpha=0.1$, as can seen in Fig.\ \ref{graph4}.
This concludes Example 6.\\
    
\subsection{Constraints with Memory}
\label{constraintswithmemory2}

When dealing with constraint functions with memory, our above derivation is
supposed to involve evaluation of an expression of the form
$$\int_{\reals^n}
\exp\left\{-\btheta\bullet\sum_{i=1}^n\bphi(x_{i-m+1}^i)\right\}p(\by|\bx)\mbox{d}\bx=
\int_{\reals^n}\prod_{i=m}^n\left[
\exp\left\{-\btheta\bullet\bphi(x_{i-m+1}^i)\right\}p(y_i|x_i)\right]\mbox{d}\bx.$$
Unfortunately, since the sliding-window kernel depends here on $y_i$, and
therefore, not fixed as before, there is no apparent single-letter characterization
of its exponential growth rate, to the best knowledge of the authors. In order to proceed, it seems necessary
to give up on some tightness and to bound the corresponding quantity using
manageable expressions. Here we only outline the starting point of the
derivation and the continuation will be deferred to future work.

We begin by applying the Jensen inequality:
\begin{eqnarray}
& &\bE\left\{\log\left[\int_{\calS_n}p(\bY|\bx)\mbox{d}\bx\right]\right\}\nonumber\\
&\le&\log\bE\left\{\int_{\calS_n}p(\bY|\bx)\mbox{d}\bx\right\}\nonumber\\
&=&\log\left[\int_{\calS_n}\frac{\mbox{d}\bx'}
{\mbox{Vol}(\calS_n)}\int_{\reals^n}\mbox{d}\by p(\by|\bx')\int_{\calS_n}p(\by|\bx)\mbox{d}\bx\right]\nonumber\\
&=&\log\left[\int_{\calS_n^2}\mbox{d}\bx\mbox{d}\bx'
\int_{\reals^n}\mbox{d}\by
p(\by|\bx)p(\by|\bx')\right]-\log\mbox{Vol}(\calS_n)\nonumber\\
&=&\log\left[\int_{\calS_n^2}\mbox{d}\bx\mbox{d}\bx'
\frac{\exp\{-\|\bx-\bx'\|^2/(4\sigma^2)}{(4\pi\sigma^2)^{n/2}}
\right]-\log\mbox{Vol}(\calS_n)\nonumber\\
&=&\log\left[\int_{\calS_n^2}\mbox{d}\bx\mbox{d}\bx'
\exp\left\{-\frac{\|\bx-\bx'\|^2}{4\sigma^2}\right\}
\right]-\log\mbox{Vol}(\calS_n)-\frac{n}{2}\log(4\pi\sigma^2)\nonumber\\
&\le&\log\bigg[\inf_{\btheta_1,\btheta_2\in\reals_+^k}\int_{\reals^{2n}}
\mbox{d}\bx_1\mbox{d}\bx_2\exp\{n(\btheta_1+\btheta_2)\bullet\bGamma-
\btheta_1\bullet\bphi(\bx_1)-\btheta_2\bullet\bphi(\bx_2)\}\times\nonumber\\
& &\exp\left\{-\frac{\|\bx_1-\bx_2\|^2}{4\sigma^2}\right\}\bigg]-
\log\mbox{Vol}(\calS_n)-\frac{n}{2}\log(4\pi\sigma^2)\nonumber\\
&=&\inf_{\btheta_1,\btheta_2\in\reals_+^k}
\bigg(n(\btheta_1+\btheta_2)\bullet\bGamma+n\log\bigg[\int_{\reals^2}\mbox{d}x_1\mbox{d}x_2
\exp\bigg\{-\btheta_1\bullet\bphi(x_1)-\btheta_2\bullet\bphi(x_2)-\nonumber\\
& &\frac{(x_1-x_2)^2}{4\sigma^2}\bigg\}\bigg]\bigg)-
\log\mbox{Vol}(\calS_n)-\frac{n}{2}\log(4\pi\sigma^2)\nonumber\\
&\dfn&n\cdot\inf_{\btheta_1,\btheta_2\in\reals_+^k}\left[(\btheta_1+\btheta_2)\bullet\bGamma+\log
Z(\btheta_1,\btheta_2)\right]-
\log\mbox{Vol}(\calS_n)-\frac{n}{2}\log(4\pi\sigma^2).
\end{eqnarray}
For a possible further improvement, one may execute a change of measures, using the
following well-known inequality for a positive random variable, $U$:
\begin{equation}
\bE_p\{\log U\}\le \log\bE_q\{U\}+D(p\|q),
\end{equation}
where $q$ is an arbitrary distribution and equality is achieved for
$q(u)=\frac{p(u)/u}{\int_0^\infty p(u')\mbox{d}u'/u'}$.
In our case,
$p(\by)=\frac{1}{\mbox{Vol}(\calS_n)}\int_{\calS_n}p(\by|\bx)\mbox{d}\bx$.
We will take the auxiliary distribution to be
$q(\by)=\frac{1}{\mbox{Vol}(\calS_n)}\int_{\calS_n}q(\by|\bx)\mbox{d}\bx$.
where $q(\by|\bx)=\calN(\bx,s^2I_n)$ (other possibilities will be considered
in the sequel). Now,
\begin{eqnarray}
& &\bE\left\{\log\left[\int_{\calS_n}p(\bY|\bx)\mbox{d}\bx\right]\right\}\nonumber\\
&\le&\log\bE_q\left\{\int_{\calS_n}p(\bY|\bx)\mbox{d}\bx\right\}+
D(p_{\bY}\|q_{\bY})\nonumber\\
&\le&\log\bE_q\left\{\int_{\calS_n}p(\bY|\bx)\mbox{d}\bx\right\}+
D(p_{\bX\bY}\|q_{\bX\bY})\nonumber\\
&\le&\log\bE_Q\left\{\int_{\calS_n}p(\bY|\bx)\mbox{d}\bx\right\}+
\frac{n}{2}\left(\frac{\sigma^2}{s^2}-\log\frac{\sigma^2}{s^2}-1\right)\nonumber\\
&=&\log\left\{\int_{\calS_n^2}\mbox{d}\bx\mbox{d}\bx'\int_{\reals^n}p(\by|\bx)q(\by|\bx')\mbox{d}\by\right\}-
\log\mbox{Vol}(\calS_n)+
\frac{n}{2}\left(\frac{\sigma^2}{s^2}-\log\frac{\sigma^2}{s^2}-1\right)\nonumber\\
&=&\log\left\{\int_{\calS_n^2}\mbox{d}\bx\mbox{d}\bx'(2\pi\sigma^2)^{-n/2}(2\pi
s^2)^{-n/2}\int_{\reals^n}\exp\left[-\frac{\|\by-\bx\|^2}{2\sigma^2}-
\frac{\|\by-\bx'\|^2}{2s^2}\right]\mbox{d}\by\right\}-\nonumber\\
& &\log\mbox{Vol}(\calS_n)+
\frac{n}{2}\left(\frac{\sigma^2}{s^2}-\log\frac{\sigma^2}{s^2}-1\right)\nonumber\\
&=&\log\left\{\int_{\calS_n^2}\mbox{d}\bx\mbox{d}\bx'[2\pi(\sigma^2+s^2)]^{-n/2}
\exp\left[-\frac{\|\bx-\bx'\|^2}{2(\sigma^2+s^2)}\right]\right\}-\nonumber\\
& &\log\mbox{Vol}(\calS_n)+
\frac{n}{2}\left(\frac{\sigma^2}{s^2}-\log\frac{\sigma^2}{s^2}-1\right)\nonumber\\
&=&\log\left\{\int_{\calS_n^2}\mbox{d}\bx\mbox{d}\bx'
\frac{n}{2}\left(\frac{\sigma^2}{s^2}-\log\frac{\sigma^2}{s^2}-1\right)\right\}\nonumber\\
&=&\log\left\{\int_{\calS_n^2}\mbox{d}\bx\mbox{d}\bx'
\exp\left[-\frac{\|\bx-\bx'\|^2}{2(\sigma^2+s^2)}\right]\right\}-\nonumber\\
& &-\frac{n}{2}\log[2\pi(\sigma^2+s^2)]-\log\mbox{Vol}(\calS_n)+
\frac{n}{2}\left(\frac{\sigma^2}{s^2}-\log\frac{\sigma^2}{s^2}-1\right).
\end{eqnarray}
Finally, the first term is handled as before, except that in the definition of
$Z(\btheta_1,\btheta_2)$, the denominator of the exponent, $4\sigma^2$, is
replaced by $2(\sigma^2+s^2)$. Somewhat more generally, if we define
$q(\by|\bx)=\calN(\alpha\bx,s^2I_n)$, we end up with
\begin{eqnarray}
& &\log\left\{\int_{\calS_n^2}\mbox{d}\bx\mbox{d}\bx'
\exp\left[-\frac{\|\bx-\alpha\bx'\|^2}{2(\sigma^2+s^2)}\right]\right\}-\nonumber\\
& &-\frac{n}{2}\log[2\pi(\sigma^2+s^2)]-\log\mbox{Vol}(\calS_n)+
\frac{n}{2}\left(\frac{\sigma^2}{s^2}-\log\frac{\sigma^2}{s^2}-1\right)+
\frac{1}{2s^2}\bE\|\bX-\alpha\bX\|^2,
\end{eqnarray}
where the last term is $\frac{n(1-\alpha)^2\bE\{X^2\}}{2s^2}$, which requires
the marginal of $X$ (e.g., the variance $\sigma_X^2$ of the truncated Gaussian
RV in Smith's model).

\section{Summary and Conclusion}
\label{conclusion}

In this work, we have addressed the classical problem of assessing the
capacity of the discrete-time Gaussian memoryless channel,
focusing on a general framework of pointwise channel input constraints, 
which are relevant in a wide spectrum of theoretical and practical scenarios,
including peak-power constraints, correlation constraints, limitations on the
relative frequency of sign changes in the channel input signal, and many others.
Our main contribution is in proposing systematic methodologies for
deriving good lower
bounds to the channel capacity in this general framework.

Two classes of lower bounds are derived based on the precise evaluation of
the asymptotic exponential behavior of the volume (that is, the volume
exponent) of the set of legitimate input vectors, namely, those
that satisfy the aforementioned constraints.
As mentioned, the mathematical analysis technique is based on extensions of the
method of types to continuous
alphabets \cite{MW25}, which can be presented also on the basis 
of the saddle-point integration method \cite{deBruijn81} and large
deviations theory \cite{DZ98}.
The first class of lower bounds applies to continuous-valued channel inputs
and relies on the
classical entropy-power inequality \cite{CT06}. The second class 
provides tighter results
at the expense of more complicated expressions to be
optimized.

The quality and generality of the bounds is demonstrated in several examples,
including the classical peak- and average power constraints, absolute value constraints,
correlation constraints, both separate and combined. Also the combination of a linear
operation and a peak power constraint is mentioned.

For future work along the same line of thought, several directions
seem to be interesting:
\begin{enumerate}
\item Further development of the second class of bounds for constraints with memory
(see the end of Subsection \ref{constraintswithmemory2}).
\item Extending the results for colored Gaussian channels.
\item Extending the results for memoryless channels that are not necessarily Gaussian.
\item Further development of the improved bounds for non-uniform inputs, in
continuation to the last part of Subsection \ref{memorylessconstraints2}.
\item Dual upper bounds for the rate-distortion function of the Gaussian
source subject to multiple constraints on the reproduction vector.
\item Applying similar techniques to the timely problem of Integrated Sensing and Communications
(ISAC) \cite{WZLSHL25}, \cite{XLCYH23}.
Here the bounds could apply to both sensing and communication: the mutual information
is relevant for users that decode the transmitted message, while those
who merely know the statistics of the transmitted signal and
attempt to estimate it, under say, the minimum mean
square error (MMSE) performance measure (considering interesting bounds related to MMSE).
\end{enumerate}

\section*{Appendix A}
\renewcommand{\theequation}{A.\arabic{equation}}
    \setcounter{equation}{0}

\noindent
{\em Proof of Lemma \ref{lemma1}.}
Define the exponential family,
\begin{equation}
f_{\btheta}(x_i)=\frac{\exp\{-\btheta\bullet\phi(x_i)\}}{Z(\btheta)},
\end{equation}
and let $f_{\btheta}(\bx)=\prod_{i=1}^nf_{\btheta}(x_i)$.
Now, for every $\btheta\in\reals_+^k$,
\begin{eqnarray}
1&\ge&\int_{\calS_n}f_{\btheta}(\bx)\mbox{d}\bx\nonumber\\
&=&\int_{\calS_n}\frac{\exp\{-\btheta\bullet\sum_{i=1}^n\bphi(x_i)\}}{[Z(\btheta)]^n}\mbox{d}\bx\nonumber\\
&\ge&\int_{\calS_n}\frac{\exp\{-n\btheta\bullet\bGamma\}}{[Z(\btheta)]^n}\mbox{d}\bx\nonumber\\
&=&\frac{\mbox{Vol}\{\calS_n\}\exp\{-n\btheta\bullet\bGamma\}}{[Z(\btheta)]^n},
\end{eqnarray}
and so,
\begin{equation}
\mbox{Vol}\{\calS_n\}\le\exp\{n\btheta\bullet\bGamma\}\cdot[Z(\btheta)]^n.
\end{equation}
Taking logarithms of both sides, dividing by $n$, and finally, passing to the
limit of $n\to\infty$, yields
\begin{equation}
v(\bGamma)\le \btheta\bullet\bGamma+\psi(\btheta),
\end{equation}
and since the left-hand side does not depend on $\btheta$, we may minimize the
right-hand side over $\btheta\in\reals_+^k$, and obtain
\begin{equation}
v(\bGamma)\le \omega(\bGamma).
\end{equation}
This completes the proof of part 1.

Moving on to part 2, select $\btheta$ such that
$\bE_{\btheta}\{\bphi(X)\}\equiv -\dot{\psi}(\btheta)=\bGamma_i$,
where $\bE_{\btheta}\{\cdot\}$ denotes expectation under $f_{\btheta}$.
This is possible
by one of the postulates of part 2. Let $c>0$ be an arbitrary constant, and consider the
set
\begin{eqnarray}
\calS_n'&=&\left\{\bx:~n\Gamma_j-c\sqrt{n}\le\sum_{i=1}^n\phi_j(x_i)\le
n\Gamma_j~~j=1,2,\ldots,k\right\}\nonumber\\
&=&\left\{\bx:~0\le\frac{1}{\sqrt{n}}\sum_{i=1}^n[\Gamma_j-\phi_j(x_i)]\le
c~~~~j=1,2,\ldots,k\right\}.
\end{eqnarray}
Now, by the multivariate version of the central limit theorem,
as $n\to\infty$, the probability of $\calS_n'$
tends to the probability of the hypercube $[0,c]^k$ under the zero-mean multivariate
Gaussian distribution with covariance matrix
$\{\mbox{Cov}_{\btheta}\{\phi_i(X),\phi_j(X)\},~i,j=1,\ldots,k\}=\ddot{\psi}(\btheta)$,
where $\mbox{Cov}_{\btheta}\{\cdot,\cdot\}$ denotes covariance under $f_{\btheta}$.
Let us denote this probability by $\Pi_c(\btheta)$. Thus, for every $\epsilon>
0$, and large enough $n$,
\begin{equation}
\label{clt}
\int_{\calS_n'}f_{\btheta}(\bx)\mbox{d}\bx\ge \Pi_c(\btheta)\cdot(1-\epsilon).
\end{equation}
On the other hand, consider the following chain of inequalities:
\begin{eqnarray}
\int_{\calS_n'}f_{\btheta}(\bx)\mbox{d}\bx&=&
\int_{\calS_n'}\frac{\exp\left\{-\btheta\bullet\sum_{i=1}^n\bphi(x_i)\right\}}{[Z(\btheta)]^n}\mbox{d}\bx\nonumber\\
&\le&\int_{\calS_n'}\frac{\exp\{-n\btheta\bullet(\bGamma-\bc/\sqrt{n})\}}{[Z(\btheta)]^n}\mbox{d}\bx\nonumber\\
&=&\frac{\mbox{Vol}\{\calS_n'\}\cdot\exp\{-n\btheta\bullet(\bGamma-\bc/\sqrt{n})\}}{[Z(\btheta)]^n}\nonumber\\
&\le&\frac{\mbox{Vol}\{\calS_n\}\cdot\exp\{-n\btheta\bullet(\bGamma-\bc/\sqrt{n})\}}{[Z(\btheta)]^n},
\end{eqnarray}
where $\bc$ is the $k$-dimensional vector whose components are equal to $c$,
and the last step is due to the fact that $\calS_n'\subseteq\calS_n$.
Combining this with eq.\ (\ref{clt}), we have
\begin{equation}
\mbox{Vol}\{\calS_n\}\ge [Z(\btheta)]^n\cdot
\exp\{n\btheta\bullet(\bGamma-\bc/\sqrt{n})\}\cdot\Pi_c(\btheta)(1-\epsilon).
\end{equation}
Taking logarithms of both sides, dividing by $n$, and passing to the limit of
$n\to\infty$, we get
\begin{equation}
v(\bGamma)\ge \psi(\btheta)+\btheta\bullet\bGamma\ge
\inf_{\btheta\in\reals_+^k}\{\psi(\btheta)+\btheta\bullet\bGamma\}=\omega(\bGamma),
\end{equation}
completing the proof of part 2.

\section*{Appendix B}
\renewcommand{\theequation}{B.\arabic{equation}}
    \setcounter{equation}{0}

\noindent
{\em Derivation of Example 6 for $q(\bx)\propto\exp\{\alpha\|\bx\|^2\}$.}

Consider Smith's model, where
\begin{equation}
q(\bx)=\left\{\begin{array}{ll}
\frac{e^{\alpha\|\bx\|^2}}{Z_n(\alpha,A)} & \bx\in\calS_n\\
0 & \mbox{elsewhere}\end{array}\right.
\end{equation}
where
\begin{equation}
Z_n(\alpha,A)=\int_{\calS_n}e^{\alpha\|\bx\|^2}\mbox{d}\bx
\end{equation}
and $\calS_n=[-A,A]^n\bigcap\calB_n(\sqrt{nP})$ with
$\calB_n(r)=\{\bx:~\|\bx\|^2\le r^2\}$.
Let us denote
\begin{equation}
J(a,b)\dfn\int_{-A}^Ae^{ax^2+bx}\mbox{d}x.
\end{equation}
The mutual information is given by
\begin{eqnarray}
I(\bX;\bY)&=&h(\bY)-\frac{n}{2}\log(2\pi e\sigma^2)\\
&=&-\bE\left\{\log\left[\int_{\calS_n}q(\bx)p(\bY|\bx)\mbox{d}\bx\right]\right\}-
\frac{n}{2}\log(2\pi e\sigma^2)\\
&=&-\bE\left\{\log\left[\int_{\calS_n}\exp\left\{\alpha\|\bx\|^2-\frac{\|\bY-\bx\|^2}{2\sigma^2}\right\}
\mbox{d}\bx\right]\right\}+\log
Z_n(\alpha,A)+\nonumber\\
& &\frac{n}{2}\log(2\pi\sigma^2)-\frac{n}{2}\log(2\pi
e\sigma^2)\nonumber\\
&=&-\bE\left\{\log\left[\int_{\calS_n}\exp\left\{\alpha\|\bx\|^2-\frac{\|\bY-\bx\|^2}{2\sigma^2}\right\}
\mbox{d}\bx\right]\right\}+\log
Z_n(\alpha,A)-\frac{n}{2}\nonumber\\
&\dfn&-A_n+B_n-\frac{n}{2}.
\end{eqnarray}
Now,
\begin{eqnarray}
B_n&=&\log Z_n(\alpha,A)\nonumber\\
&=&\log\left[\int_{\calS_n}\exp\{\alpha\|\bx\|^2\}\mbox{d}\bx\right]\nonumber\\
&=&\inf_{\theta\ge 0}\log\left[\int_{[-A,A]^n}\exp\{n\theta
P+(\alpha-\theta)\|\bx\|^2\}\mbox{d}\bx\right]\nonumber\\
&=&n\cdot\inf_{\theta\ge 0}\left\{\theta
P+\log\left(\int_{-A}^{A}e^{(\alpha-\theta)x^2}\mbox{d}x\right)\right\}\nonumber\\
&=&n\cdot\inf_{\theta\ge 0}\{\theta P+\log J(\alpha-\theta,0)\}.
\end{eqnarray}
and
\begin{eqnarray}
A_n&=&\bE\left\{\log\left[\int_{\calS_n}\exp\left\{\alpha\|\bx\|^2-
\frac{\|\bY-\bx\|^2}{2\sigma^2}\right\}\mbox{d}\bx\right]\right\}\nonumber\\
&\le&\inf_{\theta\ge 0}\left(n\theta P+
\bE\left\{\log\left[\int_{[-A,A]^n}\exp\left\{(\alpha-\theta)\|\bx\|^2-\frac{\|\bY-\bx\|^2}{2\sigma^2}\right\}
\mbox{d}\bx\right]\right\}\right)\nonumber\\
&=&\inf_{\theta\ge 0}\left(n\theta P+
\bE\left\{\log\left[\prod_{i=1}^n\int_{-A}^A\exp\left\{(\alpha-\theta)x^2-\frac{(Y_i-x)^2}{2\sigma^2}\right\}\mbox{d}x
\right]\right\}\right)\nonumber\\
&=&\inf_{\theta\ge 0}\left(n\theta P+
\sum_{i=1}^n\bE\left\{-\frac{Y_i^2}{2\sigma^2}+
\log\left[\int_{-A}^A\exp\left\{\left(\alpha-\frac{1}{2\sigma^2}-\theta\right)x^2+\frac{xY_i}{\sigma^2}\right\}\mbox{d}x
\right]\right\}\right)\nonumber\\
&=&n\cdot\inf_{\theta\ge 0}\left[\theta
P-\frac{\bE\{Y^2\}}{2\sigma^2}+\bE\left\{\log J\left(\alpha-\frac{1}{2\sigma^2}-
\theta,\frac{Y}{\sigma^2}\right)\right\}\right]\nonumber\\
&=&n\cdot\inf_{\theta\ge 0}\left[\theta
P-\frac{\bE\{X^2\}+\sigma^2}{2\sigma^2}+\bE\left\{\log J\left(\alpha-\frac{1}{2\sigma^2}-
\theta,\frac{Y}{\sigma^2}\right)\right\}\right]\nonumber\\
&=&n\cdot\inf_{\theta\ge 0}\left[\theta
P-\frac{\bE\{X^2\}}{2\sigma^2}+\bE\left\{\log J\left(\alpha-\frac{1}{2\sigma^2}-\theta,
\frac{Y}{\sigma^2}\right)\right\}\right]-\frac{n}{2},
\end{eqnarray}
where the expectations are w.r.t.\ the asymptotic marginals of the single
symbols $X$ and
$Y$, respectively. Let $\theta_\star$ denote the minimizing $\theta$ in the
definition of $B_n$. Then, $p_{X}(x)$ is given by
\begin{equation}
p_{X}(x)=\frac{e^{(\alpha-\theta_\star)x^2}\calI\{|x|\le
A\}}{J(\alpha-\theta_\star,0)}
\end{equation}
where $\theta_\star=0$ whenever
$P\ge \frac{\int_{-A}^Ax^2e^{\alpha x^2}\mbox{d}x}
{\int_{-A}^A e^{\alpha x^2}\mbox{d}x}=\frac{\partial \log
J(\alpha,0)}{\partial\alpha}$. The marginal of $Y$ is given by the
convolution between $p_{X}$ and $\calN(0,\sigma^2)$, namely,
\begin{eqnarray}
p_{Y}(y)&=&\int_{-A}^A\frac{e^{(\alpha-\theta_\star)x^2}}{J(\alpha-\theta_\star,0)}
\cdot\frac{e^{-(y-x)^2/(2\sigma^2)}}{\sqrt{2\pi\sigma^2}}\mbox{d}x\nonumber\\
&=&\frac{\exp\{-y^2/(2\sigma^2)\}}{J(\alpha-\theta_\star,0)\sqrt{2\pi\sigma^2}}
\int_{-A}^A\exp\left\{\left(\alpha-\frac{1}{2\sigma^2}-\theta_\star\right)x^2+\frac{xy}{\sigma^2}\right\}\mbox{d}x\nonumber\\
&=&\frac{\exp\{-y^2/(2\sigma^2)\}}{J(\alpha-\theta_\star,0)\sqrt{2\pi\sigma^2}}\cdot
J\left(\alpha-\frac{1}{2\sigma^2}-\theta_\star,\frac{y}{\sigma^2}\right).
\end{eqnarray}
To summarize, $C(\bGamma)\ge C_1$, where
\begin{eqnarray}
C_1&=&\inf_{\theta\ge 0}\{\theta P+\log
J(\alpha-\theta,0)\}+\frac{\int_{-A}^Ax^2e^{(\alpha-\theta_\star)x^2}\mbox{d}x}{2\sigma^2J(\alpha-\theta_\star,0)}-
\inf_{\vartheta\ge 0}\bigg\{\vartheta P+\nonumber\\
& &\int_{-\infty}^\infty\mbox{d}y 
\frac{\exp\{-y^2/(2\sigma^2)\}}{J(\alpha-\theta_\star,0)\sqrt{2\pi\sigma^2}}\cdot
J\left(\alpha-\frac{1}{2\sigma^2}-\theta_\star,\frac{y}{\sigma^2}\right)\times\nonumber\\
& &\log J\left(\alpha-\frac{1}{2\sigma^2}-\vartheta,\frac{y}{\sigma^2}\right)\bigg\}.
\end{eqnarray}
where $\theta_\star$ is the minimizing $\theta$ in the first minimization.

\end{document}